\documentclass{vgtc}                          
\ifpdf
  \pdfoutput=1\relax
  \usepackage{graphicx}
  \DeclareGraphicsExtensions{.pdf,.png,.jpg,.jpeg}
\else
  \usepackage{graphicx}
  \DeclareGraphicsExtensions{.eps}
\fi
\graphicspath{{figures/}{pictures/}{images/}{./}}
\PassOptionsToPackage{warn}{textcomp}
\usepackage{microtype}
\usepackage{textcomp}
\usepackage{mathptmx}
\usepackage{times}
\usepackage{cite}
\usepackage{tabu}
\usepackage{booktabs}
\usepackage{amsmath}
\usepackage[nolist]{acronym}
\usepackage{listings}
\usepackage[table,xcdraw]{xcolor}
\usepackage{booktabs}
\usepackage{array}
\usepackage{multirow}
\usepackage{steinmetz}
\usepackage[final]{pdfpages}
\usepackage[abbreviations]{glossaries-extra}
\usepackage{booktabs}
\usepackage{xspace}
\usepackage{color}
\usepackage{xcolor}
\usepackage{savesym}
\usepackage{setspace}
\usepackage{tikz}
\usepackage{colortbl}
\usepackage{minitoc}
\usepackage{enumitem}
\usepackage{xcolor}
\definecolor{maroon}{cmyk}{0, 0.87, 0.68, 0.32}
\definecolor{halfgray}{gray}{0.55}
\definecolor{ipython_frame}{RGB}{207, 207, 207}
\definecolor{ipython_bg}{RGB}{247, 247, 247}
\definecolor{ipython_red}{RGB}{186, 33, 33}
\definecolor{ipython_green}{RGB}{0, 128, 0}
\definecolor{ipython_cyan}{RGB}{64, 128, 128}
\definecolor{ipython_purple}{RGB}{170, 34, 255}

\usepackage{listings}
\lstdefinelanguage{iPython}{
    morekeywords={access,and,break,class,continue,def,del,elif,else,except,exec,finally,for,from,global,if,import,in,is,lambda,not,or,pass,print,raise,return,try,while},%
    morekeywords=[2]{abs,all,any,basestring,bin,bool,bytearray,callable,chr,classmethod,cmp,compile,complex,delattr,dict,dir,divmod,enumerate,eval,execfile,file,filter,float,format,frozenset,getattr,globals,hasattr,hash,help,hex,id,input,int,isinstance,issubclass,iter,len,list,locals,long,map,max,memoryview,min,next,object,oct,open,ord,pow,property,range,raw_input,reduce,reload,repr,reversed,round,set,setattr,slice,sorted,staticmethod,str,sum,super,tuple,type,unichr,unicode,vars,xrange,zip,apply,buffer,coerce,intern},%
    sensitive=true,%
    morecomment=[l]\#,%
    morestring=[b]',%
    morestring=[b]",%
    morestring=[s]{'''}{'''},
    morestring=[s]{"""}{"""},
    morestring=[s]{r'}{'},
    morestring=[s]{r"}{"},%
    morestring=[s]{r'''}{'''},%
    morestring=[s]{r"""}{"""},%
    morestring=[s]{u'}{'},
    morestring=[s]{u"}{"},%
    morestring=[s]{u'''}{'''},%
    morestring=[s]{u"""}{"""},%
    literate=
    {á}{{\'a}}1 {é}{{\'e}}1 {í}{{\'i}}1 {ó}{{\'o}}1 {ú}{{\'u}}1
    {Á}{{\'A}}1 {É}{{\'E}}1 {Í}{{\'I}}1 {Ó}{{\'O}}1 {Ú}{{\'U}}1
    {à}{{\`a}}1 {è}{{\`e}}1 {ì}{{\`i}}1 {ò}{{\`o}}1 {ù}{{\`u}}1
    {À}{{\`A}}1 {È}{{\'E}}1 {Ì}{{\`I}}1 {Ò}{{\`O}}1 {Ù}{{\`U}}1
    {ä}{{\"a}}1 {ë}{{\"e}}1 {ï}{{\"i}}1 {ö}{{\"o}}1 {ü}{{\"u}}1
    {Ä}{{\"A}}1 {Ë}{{\"E}}1 {Ï}{{\"I}}1 {Ö}{{\"O}}1 {Ü}{{\"U}}1
    {â}{{\^a}}1 {ê}{{\^e}}1 {î}{{\^i}}1 {ô}{{\^o}}1 {û}{{\^u}}1
    {Â}{{\^A}}1 {Ê}{{\^E}}1 {Î}{{\^I}}1 {Ô}{{\^O}}1 {Û}{{\^U}}1
    {œ}{{\oe}}1 {Œ}{{\OE}}1 {æ}{{\ae}}1 {Æ}{{\AE}}1 {ß}{{\ss}}1
    {ç}{{\c c}}1 {Ç}{{\c C}}1 {ø}{{\o}}1 {å}{{\r a}}1 {Å}{{\r A}}1
    {€}{{\EUR}}1 {£}{{\pounds}}1,
    literate=
    *{+}{{{\color{ipython_purple}+}}}1
    {-}{{{\color{ipython_purple}-}}}1
    {*}{{{\color{ipython_purple}$^\ast$}}}1
    {/}{{{\color{ipython_purple}/}}}1
    {^}{{{\color{ipython_purple}\^{}}}}1
    {?}{{{\color{ipython_purple}?}}}1
    {!}{{{\color{ipython_purple}!}}}1
    {\%}{{{\color{ipython_purple}\%}}}1
    {<}{{{\color{ipython_purple}<}}}1
    {>}{{{\color{ipython_purple}>}}}1
    {|}{{{\color{ipython_purple}|}}}1
    {\&}{{{\color{ipython_purple}\&}}}1
    {~}{{{\color{ipython_purple}~}}}1
    {==}{{{\color{ipython_purple}==}}}2
    {<=}{{{\color{ipython_purple}<=}}}2
    {>=}{{{\color{ipython_purple}>=}}}2
    {+=}{{{+=}}}2
    {-=}{{{-=}}}2
    {*=}{{{$^\ast$=}}}2
    {/=}{{{/=}}}2,
    commentstyle=\color{ipython_cyan}\ttfamily,
    stringstyle=\color{ipython_red}\ttfamily,
    keepspaces=true,
    showspaces=false,
    showstringspaces=false,
    rulecolor=\color{ipython_frame},
    frame=single,
    frameround={t}{t}{t}{t},
    framexleftmargin=0mm,
    numbers=left,
    numberstyle=\tiny\color{halfgray},
    backgroundcolor=\color{ipython_bg},
    basicstyle=\scriptsize\ttfamily,
    keywordstyle=\color{ipython_green}\ttfamily,
    escapechar=\¢,escapebegin=\color{ipython_green},
}

\savesymbol{spacing}
\definecolor{Orange}{rgb}{1.0, 0.5, 0.0}
\definecolor{DarkGreen}{rgb}{0, 0.5, 0.0}
\definecolor{Purple}{rgb}{0.7, 0.0, 0.7}
\definecolor{Blue}{rgb}{0.2, 0.2, 0.8}
\definecolor{Red}{rgb}{1.0, 0.0, 0.0}
\definecolor{Brown}{rgb}{0.7, 0.4, 0.1}
\definecolor{Blue}{rgb}{0, 0, 1.}
\definecolor{Green}{rgb}{0., 0.6, 0.}
\definecolor{Custom}{rgb}{0.3, 0.1, 0.2}
\definecolor{Yellow}{rgb}{0.9, 0.7, 0.0}
\definecolor{Purple}{rgb}{0.9, 0.1, 0.8}

\newcommand{\etal}{~et al.\@\xspace}

\newcommand{\eg}{e.g.\@\xspace}

\newcommand{\projectname}{\textit{HoloBeam}\@\xspace}
\newcommand{\codebase}{\textcolor{blue}{\href{https://github.com/complight/multiholo}{\textbf{GitHub:complight/multiholo}}}\@\xspace}
\newcommand{\dependbase}{\textcolor{blue}{\href{https://github.com/kaanaksit/odak}{\textbf{GitHub:kaanaksit/odak}}}\@\xspace}

\newcommand{\high}[1]{{\cellcolor{green!25} #1}}
\newcommand{\medium}[1]{\cellcolor{yellow!25} #1}
\newcommand{\low}[1]{\cellcolor{red!25} #1}

\newabbreviation{cpd}{cpd}{Cycles Per Degree}
\newabbreviation{SLM}{SLM}{Spatial Light Modulator}
\newabbreviation{HVS}{HVS}{Human Visual System}
\newabbreviation{DMD}{DMD}{Digital Micromirror Device}
\newabbreviation{AR}{AR}{Augmented Reality}
\newabbreviation{VR}{VR}{Virtual Reality}
\newabbreviation{CGH}{CGH}{Computer-Generated Holography}
\newabbreviation{FoV}{FoV}{Field Of View}
\newabbreviation{HOE}{HOE}{Holographic Optical Element}
\newabbreviation{3D}{3D}{Three-Dimensional}
\newabbreviation{CNN}{CNN}{Convolutional Neural Network}
\newabbreviation{MTF}{MTF}{Modulation Transfer Function}

\global\long\def\MTF{\gls{MTF}\xspace}
\global\long\def\CNN{\gls{CNN}\xspace}
\global\long\def\HVS{\gls{HVS}\xspace}
\global\long\def\3D{\gls{3D}\xspace}
\global\long\def\cpd{\gls{cpd}\xspace}
\global\long\def\SLM{\gls{SLM}\xspace}
\global\long\def\AR{\gls{AR}\xspace}
\global\long\def\VR{\gls{VR}\xspace}
\global\long\def\CGH{\gls{CGH}\xspace}
\global\long\def\HOE{\gls{HOE}\xspace}
\global\long\def\FoV{\gls{FoV}\xspace}
\global\long\def\DMD{\gls{DMD}\xspace}

\global\long\def\HOEs{\gls{HOE}s\xspace}

\global\long\def\SLMs{\gls{SLM}s\xspace}

\newcommand{\refFig}[1]{Fig.~\ref{fig:#1}}

\newcommand{\refTbl}[1]{Tbl.~\ref{tbl:#1}}

\onlineid{1231}
\vgtccategory{Research}
\vgtcinsertpkg
\ieeedoi{xx.xxxx/TVCG.201x.xxxxxxx}
\title{HoloBeam: Paper-Thin Near-Eye Displays}
\author{Kaan Akşit\thanks{e-mail: k.aksit@ucl.ac.uk}\\ %
        \scriptsize University College London %
\and Yuta Itoh\thanks{e-mail: yuta.itoh@iii.u-tokyo.ac.jp}\\ %
     \scriptsize The University of Tokyo %
}
\teaser{
\centering
\includegraphics[width=\linewidth]{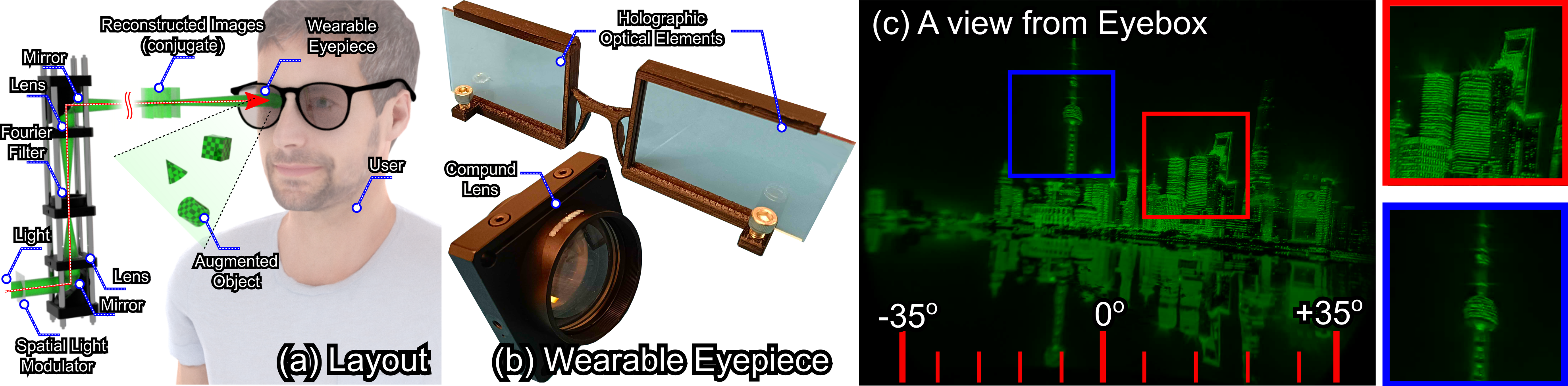}
\caption{
\projectname slim Virtual Reality headsets and Augmented Reality glasses with wide field of view and high resolutions.
(a) Layout. A light beam from a light source gets modulated by a Spatial Light Modulator (SLM).
Modulated beam reconstructs multiplane images in the vicinity of SLM (a few millimeters away).
With the help of multiple lenses forming a projection system, conjugate images are formed at considerable throw distances (e.g., a few tens of cm to a meter or two).
These conjugate images are relayed to a user's retina with the help of a slim and lightweight eyepiece.
Thus, users can perceive high-resolution \3D virtual images overlayed on their real-world scene.
(b) Wearable Eyepiece.
\projectname glasses is made of wearable eyepieces that could come in the form of a submillimeter-thick holographic optical elements or conventional lenses with a reasonable thickness.
(c) A view from eyebox.
Provided photograph demonstrates the high image quality and wide field of view of a phase-only modulating \projectname prototype when viewed through eyepieces that use conventional lenses.
Source image is from Flicker user gags9999.
}
\label{fig:teaser}
}

\abstract{
An emerging alternative to conventional Augmented Reality (AR) glasses designs, Beaming displays promise slim AR glasses free from challenging design trade-offs, including battery-related limits or computational budget-related issues.
These beaming displays remove active components such as batteries and electronics from AR glasses and move them to a projector that projects images to a user from a distance (1-2 meters), where users wear only passive optical eyepieces.
However, earlier implementations of these displays delivered poor resolutions (7 cycles per degree) without any optical focus cues and were introduced with a bulky form-factor eyepiece ($\sim50~mm$ thick).
This paper introduces a new milestone for beaming displays, which we call \projectname.
In this new design, a custom holographic projector populates a micro-volume located at some distance (1-2 meters) with multiple planes of images.
Users view magnified copies of these images from this small volume with the help of an eyepiece that is either a Holographic Optical Element (HOE) or a set of lenses.
Our \projectname prototypes demonstrate the thinnest AR glasses to date with submillimeter thickness (\eg, HOE film is only $120~\mu m$ thick).
In addition, \projectname prototypes demonstrate near retinal resolutions ($24$ cycles per degree) with a $70$ degrees-wide field of view.
}
\CCScatlist{
  \CCScatTwelve{Human-centered computing}{Visu\-al\-iza\-tion}{Visu\-al\-iza\-tion techniques}{Treemaps};
  \CCScatTwelve{Human-centered computing}{Visu\-al\-iza\-tion}{Visualization design and evaluation methods}{}
}
\begin{document}
\firstsection{Introduction}
\label{sec:introduction}

\maketitle
Emerging as the future's interface~\cite{orlosky2021telelife}, \VR and \AR glasses are expected to redefine how we interact with real and virtual environments by overlaying rendered images within our visual \FoV.
To achieve this vision, recent years have seen a strong push from the scientific community to build \AR \cite{maimone2017holographic, maimone2014pinlight} and \VR \cite{ratcliff2020thinvr,kim2022holographicglasses} glasses that are thin and lightweight.
Video see-through \AR~\cite{ebner2022video} and \VR headsets block the real-world view, whereas optical see-through \AR has no such luxury to do so.
Expectation from both \AR glasses and \VR headsets is that they should support high resolutions, wide \FoV, and optical focus cues \cite{aghasi2021optimal}.
To date, \VR headsets and AR glasses in the literature have struggled to balance these requirements, often yielding to issues related to one or more requirements (e.g., slim but low \FoV, bulky but with optical focus cues).

An emerging design alternative to conventional designs for AR glasses, Beaming Displays~\cite{itoh2021beaming}, argues that designs should separate active and passive parts in AR glasses into two discrete components.
Specifically, these Beaming Displays project images from a distance (1-2 m) to a user wearing a passive lightweight optical eyepiece.
Thanks to removing active components from AR glasses, Beaming Displays fundamentally avoid heating, computation, and power budget-related issues.
At the core, Beaming Displays aim to balance design requirements such as form-factor, resolution, \FoV, and optical focus cues.
However, the previous implementation of Beaming Displays~\cite{itoh2021beaming} provided a limited resolution with 7 \cpd and 50~mm thick bulky AR glasses.
Thus, the promises of Beaming Displays have yet to be fully realized.

\begin{table*}[ht!]
  \caption{
Comparison between \AR and \VR near-eye displays. Here, focus refers to the method used to support the optical focus cues. Optical see-through refers to the level of see-through in the real world. Wide eye-box refers to supporting varying gazes of users (above 10 mm). Moderate monocular fields of view are 20-50 degrees. Moderate resolution matches 10-20 cycles per degree. Although our work offers no mobility, it distinguishes itself as the slim and lightweight AR near-eye display, free from heating issues or limited power and computing resources.
         }
  \label{tbl:comparison}
  \begin{tabular}{m{2.9cm} m{1.2cm} m{1.0cm} m{0.6cm} m{1.0cm} m{1.0cm} m{1.43cm} m{0.82cm} m{1.3cm} m{1.1cm} m{0.8cm}}
    \toprule
	  & Focus & See-through & Eyebox & Field of View & Resolution & Form factor & Weight & Power and Compute & Heat & Mobility\\
    \midrule

          This work & 
          \high{CGH} & 
          \high{Clear} & 
          \low{Small} &
          \high{Wide} & 
          \high{High} & 
          \high{Paper-Thin} & 
          \high{Light} & 
          \high{Expandable} & 
          \high{No issue} &
          \low{Fixed}
          \\ \hline

	  Beaming Displays \cite{itoh2021beaming} & 
	  \low{Fixed} & 
	  \medium{Moderate} & 
          \high{Wide} &
	  \medium{Moderate} & 
	  \low{Low} & 
	  \low{Bulky} & 
	  \low{Regular} & 
	  \high{Expandable} & 
	  \high{No issue} &
	  \medium{Limited}
	  \\ \hline

	  Holographic VR \cite{kim2022holographicglasses}& 
	  \high{CGH} & 
	  \low{Blocks} & 
	  \low{Small} &
	  \high{Wide} & 
	  \high{High} & 
	  \medium{Thin} & 
	  \low{Regular} & 
	  \low{Limited} & 
	  \low{Issue} &
	  \high{Mobile}
	  \\ \hline

	  Video MR \cite{ebner2022video}& 
	  \high{Multiplane} & 
	  \medium{Video} & 
	  \high{Wide} &
	  \medium{Moderate} & 
	  \medium{Moderate} & 
	  \low{Bulky} & 
	  \low{Regular} & 
	  \low{Limited} & 
	  \low{Issue} &
	  \high{Mobile}
	  \\ \hline

	  Wide Étendue \cite{kuo2020high} & 
	  \high{CGH} & 
	  \medium{Moderate} & 
	  \high{Wide} &
	  \high{Wide} & 
          \low{Low} & 
	  \low{Bulky} & 
	  \low{Regular} & 
	  \low{Limited} & 
	  \low{Issue} &
	  \high{Mobile}
	  \\ \hline

	  Foveated AR \cite{kim2019foveated} & 
	  \high{Varifocal} & 
	  \medium{Moderate} & 
	  \low{Small} &
	  \high{Wide} & 
	  \high{High} & 
	  \low{Bulky} & 
	  \low{Regular} & 
	  \low{Limited} & 
	  \low{Issue} &
	  \high{Mobile}
	  \\ \hline

          Scanning Eyebox \cite{jang2018holographic} & 
	  \high{CGH} & 
	  \high{Clear} & 
	  \high{Wide} & 
	  \medium{Moderate} &
	  \medium{Moderate} & 
	  \low{Bulky} & 
	  \low{Regular} & 
	  \low{Limited} & 
	  \low{Issue} &
	  \high{Mobile}
	  \\ \hline

	  Varifocal AR \cite{akcsit2017near}& 
	  \high{Varifocal} & 
	  \medium{Moderate} & 
	  \high{Wide} &
	  \high{Wide} & 
	  \medium{Moderate} & 
	  \low{Bulky} & 
	  \low{Regular} & 
	  \low{Limited} & 
	  \low{Issue} &
	  \high{Mobile}
	  \\ \hline

	  Holographic AR \cite{maimone2017holographic} & 
	  \high{CGH} & 
	  \high{Clear} & 
	  \low{Small} &
	  \high{Wide} & 
	  \high{High} & 
	  \medium{Thin} & 
	  \low{Regular} & 
	  \low{Limited} & 
	  \low{Issue} &
	  \high{Mobile}
	  \\ \hline

          Lightfield VR \cite{huang2015light} & 
	  \high{Lightfield} & 
          \low{Blocks} & 
          \high{Wide} &
	  \medium{Moderate} & 
	  \low{Low} & 
	  \low{Bulky} & 
	  \low{Regular} & 
	  \low{Limited} & 
	  \low{Issue} &
	  \high{Mobile}
	  \\ \hline

          Pinlight Displays \cite{maimone2014pinlight} & 
	  \high{Lightfield} & 
          \medium{Moderate} & 
          \low{Small} &
	  \medium{Moderate} & 
	  \low{Low} & 
	  \medium{Thin} & 
	  \low{Regular} & 
	  \low{Limited} & 
	  \low{Issue} &
	  \high{Mobile}
	  \\ \hline

          Microlens VR \cite{lanman2013near} & 
	  \high{Lightfield} & 
          \low{Blocks} & 
          \low{Small} &
	  \low{Narrow} & 
	  \low{Low} & 
	  \medium{Thin} & 
	  \low{Regular} & 
	  \low{Limited} & 
	  \low{Issue} &
	  \high{Mobile}
	  \\

    \bottomrule
  \end{tabular}
\vspace{-3mm}
\end{table*}

This paper introduces a new Beaming Display named \projectname following recent \CGH algoritmic methods and holographic eyepiece designs.
In \projectname, a holographic projector reconstructs multiplanar images at a target location away from that projector (1-2 m).
As the reconstructed images are small, a user wearing an eyepiece made from a \HOE perceives a magnified version of the reconstructed multiplanar images.
Unlike previous implementations of Beaming Displays, \projectname delivers slim AR glasses with high resolutions (24 \cpd) and wide \FoV (70 degrees).
However, this exploration in improving Beaming Displays with \projectname comes at the cost of users being fixated in front of the projector, where Beaming Displays provide room for mobility.
Specifically, our work introduces the following contributions:

\begin{itemize}[leftmargin=3mm]

\item \textbf{Thin Eyepiece Designs for \projectname.}
We provide a theoretical analysis of our slim \HOE-based eyepiece design (submillimeter).
We verify our findings by building an in-house \HOE recording hardware and demonstrate our \HOEs in an actual \projectname prototype.

\item \textbf{Learned \CGH pipeline for \projectname.}
We introduce a learned \CGH pipeline that can calculate multiplanar holograms for our display prototypes at interactive rates (24 ms per frame).
Uniquely, This learned \CGH pipeline generates multiplanar holograms without using depth maps.
Our method is the first to attempt a learned 3D hologram generation pipeline with RGB inputs but not RGBD.

\item \textbf{\projectname Display Prototypes.}
We instantiate two experimental display prototypes to demonstrate our outcomes.
In the phase-only prototype, we use a 4k phase-only \SLM and a conventional magnifier glass as an eyepiece.
In the amplitude-only prototype, we show a cost-effective alternative to this prototype using amplitude-only \SLM s and a \HOE as an eyepiece.

\end{itemize}

Our current \projectname prototypes provide a limited eye-box.
At this time, \projectname focuses on improving form factor, weight, \FoV, and image quality, and our implementation do not offer users the freedom to move in front of a projector.
Thus, user mobility and eye-box in our approach stand as two key research issues that we aim to deal with in the future.
We believe our work could positively impact \AR use cases in automotive and office work applications in their current form.
Our code is also available at \codebase.

\section{Related Work}
\label{sec:related_work}
Our work introduces a new holographic \VR/\AR glasses design paving the way for devices that ultimately appear to be ordinary glasses.
Hence, we review relevant literature on \AR Glasses, the use of \HOEs in \AR Glasses, and algorithmic approaches in \CGH.

\subsection{Augmented Reality Glasses}
Existing \AR glasses can be broadly categorized as conventional and holographic approaches.
In both approaches, images generated by a light engine equipped with an \SLM are relayed to a user's retina using an eyepiece.
Curious readers can consult the work by Koulieris \etal \cite{koulieris2019near} for a detailed survey on design approaches.
We also suggest our readers review \refTbl{comparison} as they read through this section.

\paragraph{Conventional \AR Glasses.}
In our definition, conventional \AR glasses rely on static components to relay an incoherent image source to a user's retina.
Most common types of these \AR glasses rely on an optical relay called bird-bath optics~\cite{cameron2009application,dunn2017wide}, a combination of a beam-splitter and a beam-combiner.
Freeform optics~\cite{bauer2014two} can help improve the optical qualities of these conventional \AR glasses in terms of resolution, \FoV, and form factor.
On the other hand, folding optical paths using prism-like waveguides \cite{hu2014high, cakmakci2004compact} or diffractive waveguides \cite{cameron2009application} could also help improve the form factors of \AR glasses.
However, either freeform \AR glasses that encapsulate unique diffractive components~\cite{nikolov2021metaform} or other waveguide-based \AR glasses are typically fixed focus or often times do not converge to reliable benefits as they often arrive with tradeoffs in \FoV, form-factor, and resolution.
There are also variants of conventional \AR glasses that can provide near-accurate optical focus cues~\cite{koulieris2019near}.
Such \VR headsets or \AR glasses with optical focus cues are broadly categorized as varifocal \cite{akcsit2017near,hamasaki2019varifocal,kim2019foveated}, multiplane \cite{narain2015optimal, tan2018polarization,lee2019tomographic,cui2017optical,ebner2022video}, focal surface \cite{hiroi2021focal,matsuda2017focal,akcsit2019manufacturing}, focus-invariant \cite{konrad2017accommodation} and lightfield \cite{lanman2013near,maimone2014pinlight,akcsit2015slim,huang2015light,zhan2018high,song2019design}.
Although all these types offer exciting solutions, to our knowledge, they largely suffer from problems related to weight, bulk, power, resolution, form factor, or heat.

\begin{figure*}[!ht]
\centering
\includegraphics[width=0.9\textwidth]{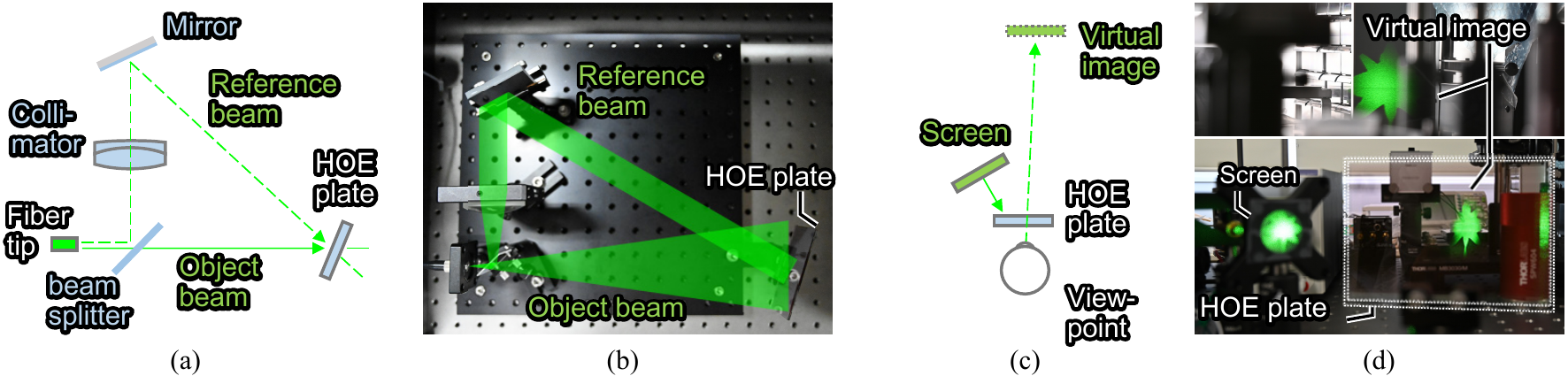}\vspace{-3mm}
\caption{
Our \HOE recording setup and functional test of a recorded \HOE lens. 
(a) A schematic diagram of our setup to record a transmissive \HOE lens. 
(b) A birds-eye view of the actual recording optics. 
(c) A schematic diagram of the test, (d) showing the images from the viewpoint.}
\label{fig:hoe_fabrication_and_test}
\vspace{-3mm}
\end{figure*}

\paragraph{Holographic \AR Glasses.}
Unlike traditional focus supporting \AR glasses, holographic \AR glasses can manipulate both phase and amplitude of light, promising potential improvements in light efficiency, resolution, and dynamic range.
Holographic \AR glasses can offer sunglasses-like form factors \HOEs \cite{maimone2017holographic} due to the use of thin diffractive optics.
These holographic \AR glasses can also offer high resolutions and support optical focus cues with continuous depth representations \cite{maimone2017holographic}.
However, the eye-box of holographic \AR glasses is typically limited due to the limited bandwidths of existing \SLMs (étendue).
To improve the eye-box of holographic \AR displays, researchers explored opportunities in scanning a light source over an \SLM or using multiple light sources \cite{jang2018holographic, hedili2019light}. 
Unfortunately, these solutions arrive at the cost of hardware complexity and additional bulk.

Most recently, people have explored structured diffusers \cite{kuo2020high} and learned phase masks \cite{baek2021neural} to replace HOEs and expand the eye-box.
But, these solutions typically arrive with image quality-related issues and require precision alignment in manufacturing.
\projectname follows common holographic \AR glasses that uses HOEs and shares similar shortcomings, namely eye-box-related issues.
Unlike previous holographic \AR glasses, \projectname offers significant improvements in fixing issues related to weight, form factor, power, and computation.
Beyond our review, we refer our readers to a survey by Chang \etal \cite{chang2020toward} on holographic \AR glasses.
To our knowledge, \projectname is the first holographic thin \AR glasses that offers a new candidate to resolve issues related to form factor, resolution, \FoV, weight, and power.

\subsection{Thin Relay Optics for Augmented Reality Glasses}
With thickness not exceeding a millimeter, \HOEs are considered as thick diffraction gratings (thicker than a few multiples of visible light wavelengths) \cite{close1975holographic}.
These wavelength-selective \HOEs interfere efficiently with beams that match angles of incidence in their recording.
Optical beams not meeting these conditions will pass through recorded \HOEs leading to maintaining high transparency in seeing the real world through \AR glasses.
Therefore, \HOEs can play a central role in \AR glasses designs as one of the technologies to realize compact and lightweight optical beam combiners~\cite{ando1999head, chang2020toward}.
The manufacturing methods of \HOEs can be broadly classified into analog and digital methods.
The former reproduce replicas of constitent actual lens systems \cite{sweatt1977describing}.
In contrast, the latter reproduce a computationally designed optical system by controlling amplitude or phase modulation patterns programmatically in \HOE recording \cite{jang2020design}.
\projectname follows \HOE design and manufacturing methods in analog \HOE recording.

\subsection{Computer-Generated Holography}
\CGH comprises a family of methods that approach the hologram calculation problem with algorithmic strategies \cite{brown1969computer,yatagai1974three}.
Our readers could find an extensive survey of modern \CGH methods in the work by Pi \etal \cite{pi2022review}.
The scene representation methods used at each \CGH algorithm could help the categorization of \CGH algorithms.
Broadly, \CGH can rely on scene representations based on the light field \cite{shi2017near,hamann2018time,jang2018holographic}, multiplane \cite{shi2021towards, symeonidou2018colour,kavakli2022realistic}, and point clouds \cite{chen2009computer,maimone2017holographic}.
Alternative scene representations such as polygon based ones \cite{im2014phase} are also available in the literature.
There are also learned algorithmic methods that can help improve speed, accuracy, and image quality in CGH \cite{zhang20173d,chakravarthula2020learned,choi2021neural,kavakli2022learned}.
In our work, we follow multiplane scene representations, and we use gradient-based optimization methods used in many earlier works to generate a dataset of holograms \cite{zhang20173d}.
Using this dataset, we derive a new learned-\CGH pipeline.
Unlike previous learned-\CGH methods, our \CGH optimization pipeline does not require depth information of a scene in hologram calculation and decides on the targets by itself.
Our \CGH pipeline runs at interactive rates on an average compute resource.

\section{\projectname : Holographic Beaming Displays}
\label{sec:holobeamer}
\projectname requires three primary components for a complete display system.
These components are a holographic projector, an eyepiece, and a software pipeline to help calculate \3D multiplanar holograms.
We also provide the layout of \projectname in \refFig{teaser}.

\subsection{Holographic Projector}
\label{subsec:holographic_projector}
Like other display systems, \projectname requires a light engine that contains light sources and an \SLM.
Given that \projectname is a holographic projector, it requires a coherent or partially coherent light source.
In the meantime, the \SLM of \projectname, could either modulate phase or amplitude of light or both - full complex.
As depicted in the basic layout \refFig{teaser}, the images generated using these \SLMs could be imaged to the desired location using a set of lenses, forming 4f or 2f imaging systems or a more advanced multi-lens system.
The conventional theoretical limits of Beaming Displays are covered in the original paper of Beaming Displays \cite{itoh2021beaming}.
However, it should be noted that the limits of resolution in holographic approaches are an open scientific debate \cite{kozacki2010resolution}, which we also agree.

\subsection{Eyepiece}\label{subsec:eyepiece}

Our holographic projector modulates light to reproduce the desired light field near the user's viewpoint.
Since this light field is so close to the eye, the eye cannot focus on it.
Besides standard optical components (\eg, lenses or mirrors), we can use \HOE to convert this light field to images on the retina that users can accommodate and view.

\paragraph{Transmissive HOEs.}
We base our work on transmissive HOEs using photopolymer films as the material (see Xiong\etal \cite{xiong2021holographic} for more).
When the photopolymer film is exposed to light, the monomers polymerize according to the interference fringes, creating an unevenly distributed polymer structure.
Because of the difference in refractive index between the monomer and polymer, this structure behaves as a phase hologram.
If the thickness of the film is sufficiently thicker than the wavelength of the light, we can treat it as a volume hologram.
Each diffraction grating structure formed within a volume hologram reflects a portion of the incident light.
At a given angle of incidence, if the optical path difference satisfies an integer multiple of the wavelength, all reflected light is intensified. 
In this case, the incident light is reflected almost entirely at one specific angle. 
This angle is called the Bragg angle, and the case where the incident light satisfies the Bragg angle is called the Bragg condition.

\begin{figure}[tb]
\centering
\includegraphics[width=0.9\columnwidth]{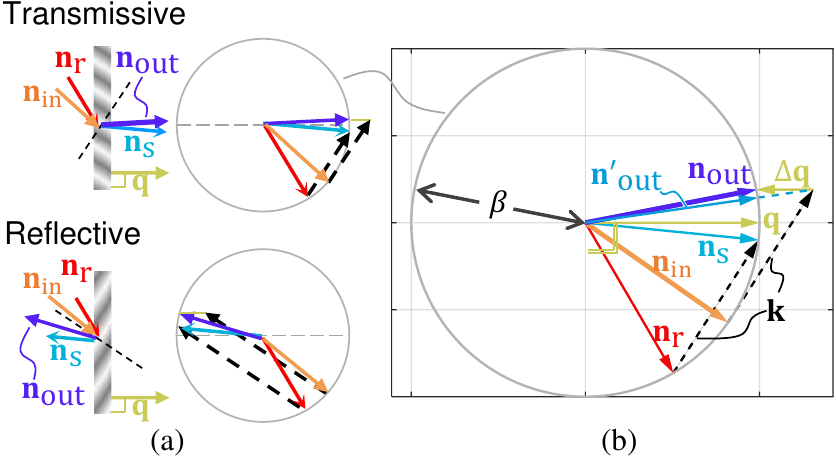}\vspace{-2mm}
\caption{Describing the principle of volume holograms based on the Bragg condition. (a top) A transimissive case. (a bottom) A reflective case. (b) Relationship of wavenumber vectors as described in Sec.~\ref{subsec:eyepiece}. Note that this setup is a special case when the wavelength of the light at the printing process of the volume holograms and that of the reconstruction are the same.
}
\label{fig:bragg-diffraction}
\end{figure}

Based on this principle, diffractive optical elements using volume holograms are selective in wavelength and angle of incidence.
Often the case, practically, it is sufficient to consider only the first-order reflected light in the numerical analysis of the reflection direction.
If the incident light is sufficiently close to the Bragg angle and the wavelength at the hologram exposure, we can approximate its behavior well with Kogelnik's coupled wave theory~\cite{prijatelj2013far}.
In this theory, the k-vector closure method (KVCM) gives the ray behavior of the incident light (\refFig{bragg-diffraction}).


\global\long\def\sinc{\operatornamewithlimits{sinc}} 
\global\long\def\Vector#1{\mathbf{\MakeLowercase{#1}}}
\global\long\def\normal{\Vector{n}}
\global\long\def\nReference{\normal_{\mathrm{r}}}
\global\long\def\nSignal{\normal_{\mathrm{s}}}
\global\long\def\kVector{\Vector{k}}
\global\long\def\nIn{\normal_{\mathrm{in}}}
\global\long\def\nOut{\normal'_{\mathrm{out}}}
\global\long\def\nOutCorrected{\normal_{\mathrm{out}}}
\global\long\def\qVector{\Vector{q}}
\global\long\def\deltaQVector{\Delta\qVector}
\global\long\def\diffractionEfficiency{\eta}
\global\long\def\betaValue{\beta}
\global\long\def\cRerefence{c_\mathrm{R}}
\global\long\def\cSignal{c_\mathrm{S}}
\global\long\def\refractiveIndex{n}
\global\long\def\refractiveIndexAverage{\refractiveIndex_0}
\global\long\def\refractiveIndexModulation{\refractiveIndex_1}
\global\long\def\wavelength{\lambda}

\global\long\def\Transpose{\mathrm{T}}
\global\long\def\incidentAngle{\theta}
\global\long\def\hologramThickness{d}

\global\long\def\asin{\operatornamewithlimits{asin}} 
\global\long\def\pixelpitch{p}


\paragraph{Bragg Diffraction Analysis}
From here, we briefly explain the theory of Bragg diffraction~ \cite{kogelnik1969coupled, fally2012experimental, prijatelj2013far, kim2017holographic}.
\refFig{bragg-diffraction}(b) describes the behavior of incident light in a transmission volume hologram. In this explanation, we also assume the wave vectors are defined inside the photopolymer medium. In real use cases, we have to consider the refraction and reflection at the boundary between the air and the photopolymer plate that consists of several layers, including a protection layer.

Let $\nReference$ be the wave vector of the reference beam incident on the photopolymer plate when creating a volume hologram, and $\nSignal$ be the wave vector of the signal beam incident from the opposite side of $\nReference$. Then, assuming that the two vectors have the same wavelength $\wavelength$, their lengths or the wave numbers are given as a constant $\betaValue=2\pi\refractiveIndexAverage/\wavelength$, where $\refractiveIndexAverage$ is the average refractive index of the photopolymer.
The two incident beams interfere with each other and create a volume hologram.

We define the k vector $\kVector$ as the wave vector extending in the direction of the interference fringes created by these two rays. These three vectors then satisfy:
\begin{equation}
\nSignal=\nReference+\kVector.
\end{equation}
Next, let us use this volume hologram to reproduce light.
Let $\nIn$ be the wave vector of the incident light.
Then, from the Bragg condition above, we can calculate the reconstructed light $\nOut$ as follows:
\begin{equation}
\nOut=\nIn+\kVector.
\end{equation}
In general, however, $\nIn$ does not align with $\nReference$. Due to this discrepancy, the naive prediction result $\nOut$ is known to deviate from the observation and needs to be corrected:
\begin{equation}
\nOutCorrected=\nIn+\kVector+\deltaQVector,
\end{equation}
where $\deltaQVector$ is a vector with the direction same as the surface normal. Its length is calculated so that $\nOutCorrected$ is placed on the circle with radius $\betaValue$. We  define a vector $\qVector$ perpendicular to the surface with length $\betaValue$.

The above equations only determine the output beam's direction. We are also interested in how much incoming light gets diffracted. The theory gives the diffraction efficiency $\diffractionEfficiency$ as follows~\cite{kogelnik1969coupled, fally2012experimental}:
\begin{align}
\diffractionEfficiency&=\left[ \nu \sinc\left(\sqrt{\nu^2+\xi^2}\right)\right]^2, \label{eq:diffractionEfficiency}\\
\nu &= \frac{\pi\refractiveIndexModulation\hologramThickness}{\wavelength\sqrt{\cRerefence\cSignal}},\ \ \ \xi = \frac{||\deltaQVector||\hologramThickness}{2\cSignal},\\
\cRerefence &= \nIn^\Transpose\qVector/\betaValue^2 (= \cos\incidentAngle),\ \ \ \cSignal =  \nOutCorrected^\Transpose\qVector/\betaValue^2,
\end{align} 
where $\cRerefence$ and $\cSignal$ are called the obliquity factors \cite{kogelnik1969coupled}.
Note that $\incidentAngle$ is the incident angle of $\nIn$, $\refractiveIndexModulation$ is the refractive index modulation of the photopolymer, and $\hologramThickness$ is the photopolymer thickness (See also eq.~(42, 43) in \cite{kogelnik1969coupled}).

As a special case, if there is no slant, i.e. the gratings are aligned with the photopolymer normal, and  the input beam is ideal, i.e., $\nIn=\nReference$, then $\xi=0$ and $\cRerefence=\cSignal$. This simplifies the diffraction efficiency to the following as given as eq.~(45) in \cite{kogelnik1969coupled},
\begin{equation}
\diffractionEfficiency=\left[\nu \sinc(\nu)\right]^2=\sin^2\nu=\sin^2\left(\frac{\pi\hologramThickness\refractiveIndexModulation}{\wavelength\cos\incidentAngle}\right).
\end{equation}

\begin{figure}[tb]
\includegraphics[width=0.45\columnwidth]{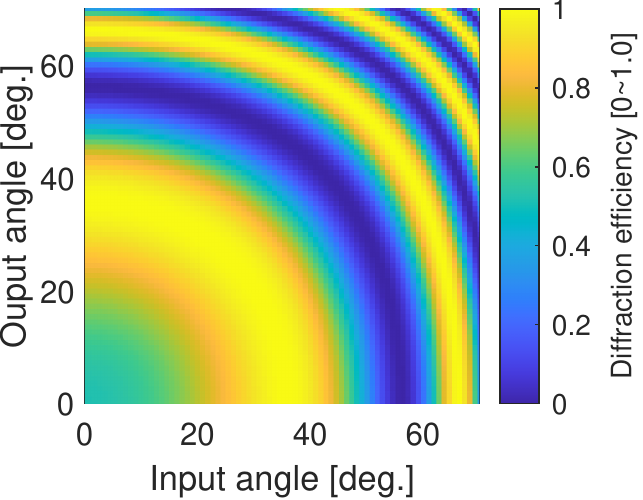}
\includegraphics[width=0.45\columnwidth]{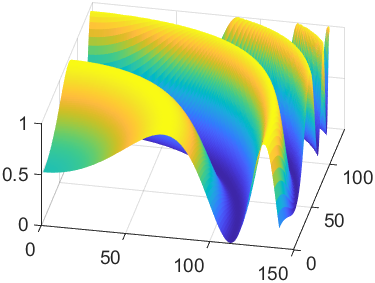}\vspace{-2mm}
\caption{Diffraction efficiency simulation of a transmissive volume hologram. (left) 2D visualization. (right) 3D visualization.}
\label{fig:diffraction_examples}
\vspace{-2mm}
\end{figure}

Figure~\ref{fig:diffraction_examples} showcases a simulation of volume hologram for Bragg conditions. 
In this simulation, we set the wavelength $\wavelength = 532$ nm, the grating thickness $\hologramThickness=30$[$\mu$m], and the maximum refractive index modulation $\refractiveIndexModulation=0.04$. The parameters are intended to replicate a commercial photopolymer Beyfol HX120 from Covestro AG. The figure shows when the reference and the signal beams take incident angle pairs between 0 to 70 degrees. One can see that there are angle pairs that give $\diffractionEfficiency~=1$, meaning such a condition can direct the input beam to the output direction with almost 100\% efficiency.

\subsection{Learned Computer-Generated Holography Pipeline}
Our learned \CGH pipeline builds upon state-of-the-art optimization methodology of hologram generation for dataset creation~\cite{curtis2021dcgh, lee2022high, kavakli2022realistic,choi2021neural,kim2022holographicglasses,zhang20173d, shi2022end}.
Firstly, we introduce this optimization method.
Then, we will explain how our learned method differs from these methods and their learned derivatives.
To our knowledge, our learned method is the first attempt towards generating \3D holograms only using RGB inputs but not RGBD inputs.

\subsubsection{Hologram Dataset Generation}
A \projectname projector generates multiplane images at the desired projection distance by optically relaying images generated close to an \SLM. 
The optical relay here refers to the set of lenses used in front of an \SLM as depicted in \refFig{teaser} layout.
As ideally, the optics take care of relaying operation in \projectname, the remaining computational challenge is calculating the ideal hologram pattern, $O_h$, for a phase-only \SLM to generate \3D images at various image planes $Z_0, Z_1, ..., Z_n$ in close vicinity of a \SLM. 
This calculation could be achieved by propagating a hologram pattern, $O_h$, using a light transport model.
Such a model could simply be written as a convolution operation \cite{sypek1995light},
\begin{equation}
u(x,y,z)=O_h(x,y) * h(x,y,z),
\end{equation}
where $u$ represents the complex amplitude at a target image plane, $z$ represents the distance between a hologram and a target image plane, and $h$ represents a convolutional kernel that simulates light transport.
Note that $h$ could also be learned from actual hardware using camera captures \cite{kavakli2022learned}, or could be replaced with a \CNN \cite{choi2021neural} using these captures.
The resultant intensity, $|u|^2$, could then be compared against a target image, $T$, using a loss function, $\mathcal{L}(|u|^2, T)$.
Over a few successions of iterations, a hologram could be generated from target images dedicated to each plane ($T_0, T_1,..., T_n$).
These target images are typically generated using an image or a photograph and their corresponding depth maps (\eg, \cite{kavakli2022realistic,choi2021neural,kim2022holographicglasses,zhang20173d}).

\subsubsection{Our Learned Method}
Our learned \CGH pipeline aims to produce multiplane holograms \textit{without requiring the scene's depth information}.
Typical and conventional scenes or images often do not arrive with depth information.
Although their depth information could be estimated reliably using a \CNN \cite{alhashim2018high}, given the wide availability of 2D images and photographs, we see value in easing a user's workflow in generating holograms without access to the depth information.
This way, in the future, existing 2D digital content (\eg, games, movies) could potentially be converted to \3D holograms without having two go through multiple steps of estimating depth and generating holograms.

The optimizations discussed previously \cite{kavakli2022realistic,choi2021neural,kim2022holographicglasses,zhang20173d} could be used to generate a hologram dataset from RGBD data.
A hologram generation model could then be trained using RGB images as input while discarding their depth channel and their corresponding optimized holograms.
Such a model could be trained and tested using the forward model found in Listings~\ref{list:forward_model}.
In this forward model, a \CNN could estimate an output with two channels from a single color of an input image.

In the meantime, we should note that holograms generating images in proximity to an \SLM typically generated using the Double-Phase coding method \cite{hsueh1978computer,maimone2017holographic,shi2017near}.
Inspired by Double-Phase coding, these estimated output channels could then be compiled into a phase-only hologram pattern following a checkerboard-like pattern, maintaining the look of a Double-Phase coded hologram.

\begin{figure}[tb]
\begin{center}
\begin{minipage}{.43\textwidth}
\begin{lstlisting}[language=iPython,escapeinside={(*}{*)},caption={
The learned differentiable model used in estimating multiplane phase-only holograms (Pythonic abstraction).
This routine runs for every color primary separately. 
},label=list:forward_model]
def estimate(x, n):
    """
    Parameters
    ----------
    x : torch.tensor
        Single color images [kx1xmxn].
    n : torch.nn.modulelist
        Network model used in training.
    
    Returns
    -------
    (*\textbf{$O_h$}*) : torch.tensor 
         Estimated holograms [kx1xmxn].
    """
    y = n.forward(x)
    a = y[:, 0]
    b = y[:, 1]
    (*\textbf{$\phi$}*)[:, :, 0::2, 0::2] = a[:, :, 0::2, 0::2]
    (*\textbf{$\phi$}*)[:, :, 1::2, 1::2] = a[:, :, 1::2, 1::2]
    (*\textbf{$\phi$}*)[:, :, 0::2, 1::2] = b[:, :, 0::2, 1::2]
    (*\textbf{$\phi$}*)[:, :, 1::2, 0::2] = b[:, :, 1::2, 0::2]
    (*\textbf{$\phi \rightarrow O_h$}*)
    return (*\textbf{$O_h$}*)
\end{lstlisting}
\end{minipage}
\end{center}
\vspace{-12mm}
\end{figure}

\section{Implementation}
This section will detail the making of \HOE in our prototypes, our two prototypes used in our evaluations, and our learned \CGH pipeline.

\subsection{\projectname Thin Eyepiece}
We used photopolymer sheets as the recording material for our \HOE.
Specifically, we used holographic film from Litiholo, a 2x3 inch plate consisting of a photopolymer applied to a 2.0 mm glass plate. 
The photopolymer consists of a 60-micron tri-acetyl cellulose (TAC) film substrate, a 60-micron recording layer, and a protective film layer.
The total thickness, including the glass substrate, was 2.28 mm.
In \HOE recording, the reference and object beams hit the plate simultaneously.
If these beams reach the same side of the plate, we obtain a transmissive \HOE; if they reach from both sides, we obtain a reflective \HOE.

In general, according to the principle of Bragg diffraction, a transmissive \HOE requires a shallower diffraction structure than a reflective \HOE.
In other words, the design tolerance of the diffraction structure is larger for transmission \HOEs, and they are more resistant to disruptions such as physical vibrations during fabrication.
Therefore, we adopted the transmissive design in this study.
For ordinary AR display applications, reflective \HOEs are more often used due to the advantages of stray light prevention, high transmission efficiency, and placement of the built-in display.

\begin{figure}[tb]
\includegraphics[width=0.9\columnwidth]{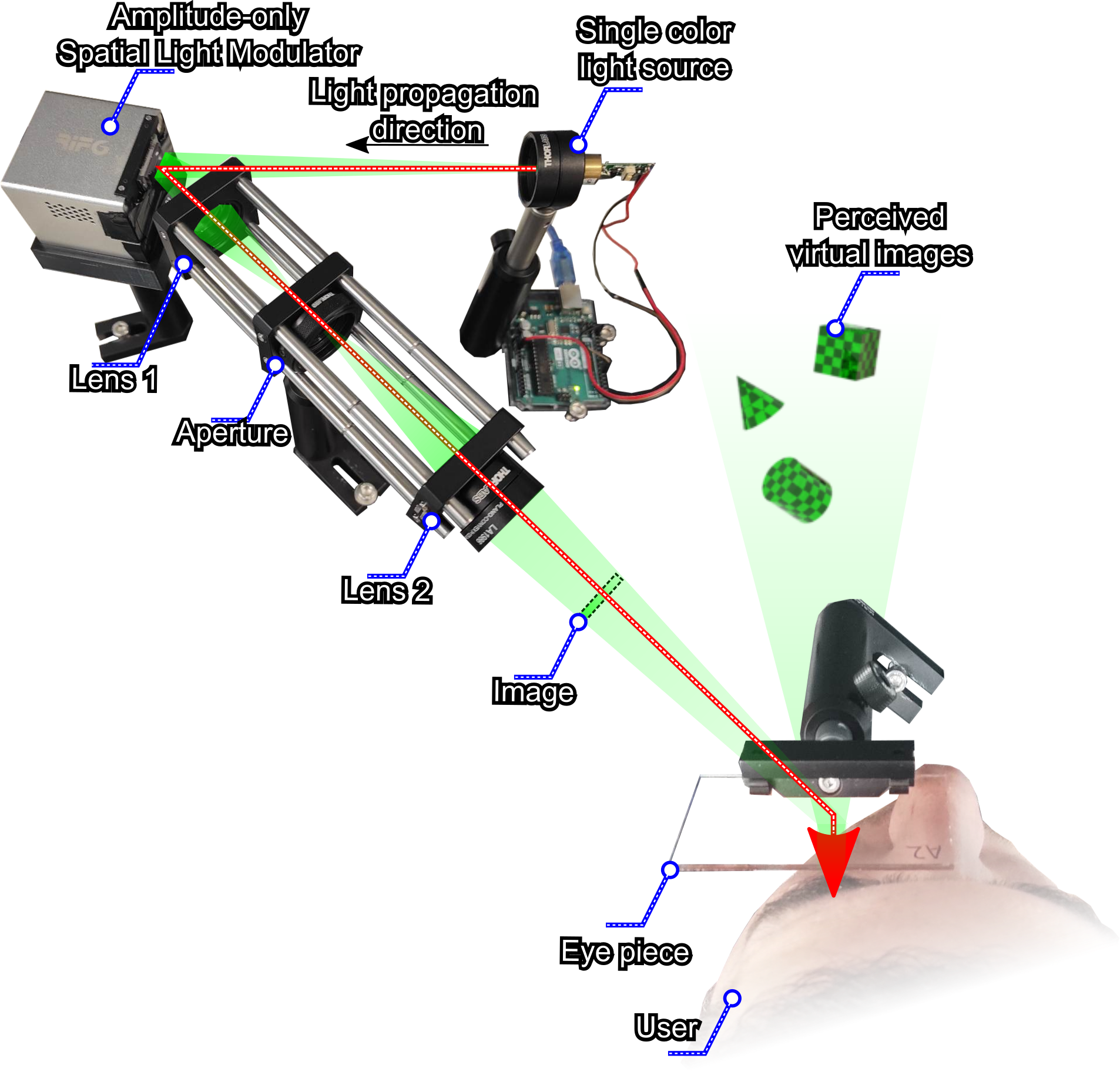}\vspace{-2mm}
\caption{
Amplitude-only \projectname prototype. 
Here, a single-color laser illuminates an amplitude-modulating \SLM.
The generated image on the \SLM is relayed to an image plane.
\HOE eyepiece relays the image on that image plane to a user’s retina, and the user perceives fixed-focus virtual images.
The distance between the projector assembly and the eyepiece is 30 cm and not drawn at true scale to avoid a wider figure.
}
\label{fig:amplitude_only_hardware}
\vspace{-3mm}
\end{figure}

\subsection{\projectname Prototypes}
\paragraph{Amplitude-only \projectname}
Our amplitude-only prototype is built to demonstrate how compact, slim, and cost-effective our solution could be.
This prototype uses a green laser diode with a 532 nm wavelength and 10 mW optical power.
We harvested a laser diode from a generic laser pointer and drove it with an IRF540N MOSFET and an Arduino microcontroller.
We collimate this coherent light with a 100 mm focal length lens, Thorlabs LA1509.
The collimated beam then arrives at an amplitude-only \SLM, specifically a \DMD with 854 by 480 pixels and $5.4 \mu m$ pixel pitch.
This specific \DMD is harvested from a pico projector, RIF 6 Cube, and we can only push 8-bit frames without any access to timing or individual binary frames.

\begin{figure}[tb]
\includegraphics[width=0.9\columnwidth]{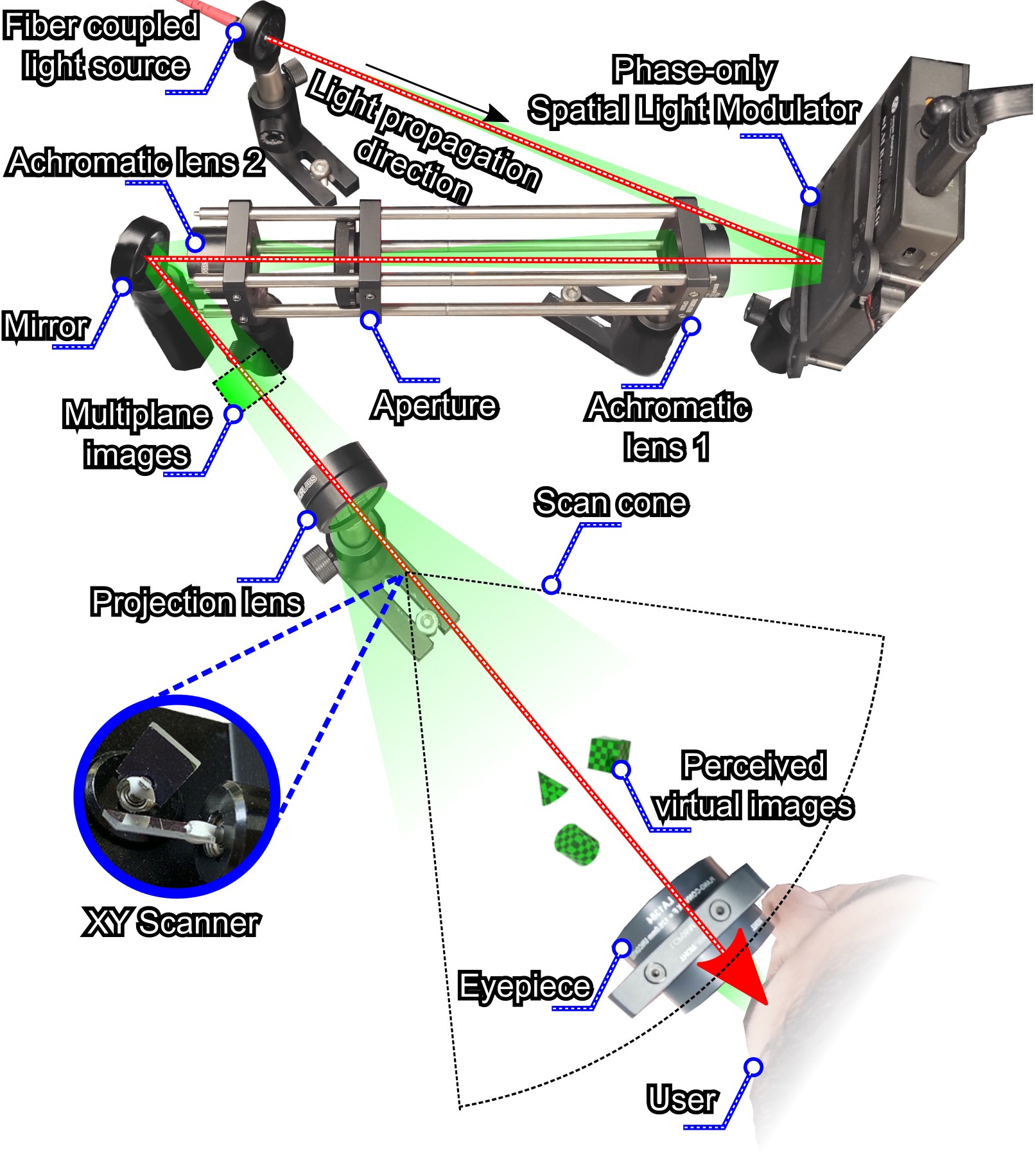}\vspace{-2mm}
\caption{
Phase-only \projectname prototype.
A fiber-coupled multicolor light source illuminates a phase-only \SLM.
Multiplane images generated from this \SLM are filtered with a pinhole using an aperture and a 4f imaging system.
These images are then projected at a meter distance using a 2f imaging system as a projection lens.
Note that an XY scanner follows the projection lens to steer the beam toward a user.
A user wearing an eyepiece composed of a cascade of lenses perceives images with multiple focuses.
The distance between the projector assembly and the eyepiece is a meter in reality and not drawn at the actual scale to avoid a wider figure.
}
\label{fig:phase_only_hardware}
\vspace{-2mm}
\end{figure}

\begin{figure}[tb]
\includegraphics[width=0.9\columnwidth]{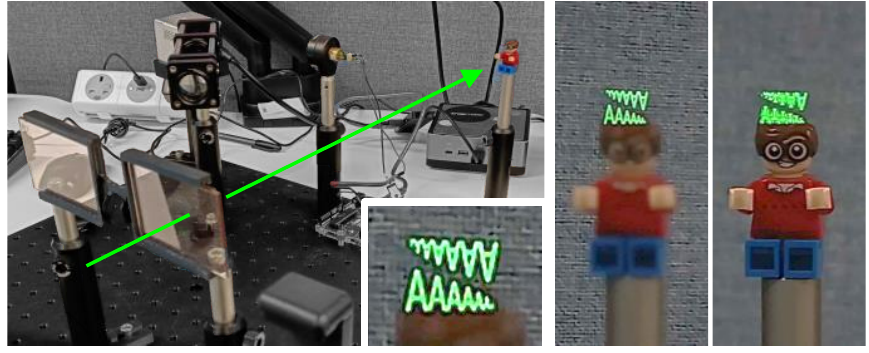}\vspace{-2mm}
\caption{
An actual photograph of a see-through image from our amplitude-only \projectname prototype.
The insets demonstrate how the generated virtual image is focused at a far plane (1.5 meters away). 
}
\label{fig:amplitude_test}
\vspace{-3mm}
\end{figure}

The modulated beam from \DMD is relayed to an image plane with a throw distance of 30 cm using 4f optics.
Following the path from \DMD towards our image plane, we use a 50 mm plano-convex lens, Thorlabs LA1131, an adjustable aperture at the Fourier plane, Thorlabs SM1D2D, and a 150 mm bi-convex lens, LB1437.
This prototype has no beam steering capability and uses a fixated \HOE eyepiece at a location 30 cm away from the projection assembly.
\refFig{amplitude_only_hardware} shows the entire assembly of our amplitude-only \projectname prototype, and \refFig{amplitude_test} shows an example see-through capture. 
As demonstrated in our supplementary materials, we could also build a vertical-layout assembly.

\paragraph{Phase-only \projectname}
We constructed our phase-only \projectname prototype to emonstrate a system with high image quality and a wide \FoV (see \refFig{phase_only_hardware}).
The prototype uses fiber-coupled light sources (420, 520, and 638 nm - Fisba ReadyBeam) and is also equipped with an incoherent RGB LED light source, which can be activated as needed. 
The phase-only \SLM in our prototype is a Jasper Display \SLM Research kit (2400 by 4094 pixels and $3.74 \mu m$ pixel pitch).
The modulated beam from the \SLM generates multiplanar images after passing through a 4f system and reflecting off a mirror used for path folding (Thorlabs ME1-P01). 
From the \SLM to the path folding mirror, we used the following optical components: an achromatic lens (Thorlabs AC254-100A-ML), an aperture located at the Fourier plane (Thorlabs SM1D12D), and another achromatic lens (Thorlabs AC254-050-A-ML).

We project the reconstructed multiplane image to an eyepiece using a 2f system consisting of Thorlabs AC254-150-A-ML.
Although there is an optional galvanometer scanner (AT20-2278) available, we did not utilize it in this work. 
At a meter throw distance, our user perceives multiplane images by looking through our eyepiece composed of two lens cascades.
From the projection toward the user, these lenses in our eyepiece are Thorlabs LA1384 and Thorlabs LA1050.
For more details, please consult our supplementary.

\begin{figure}[tb]
\centering
\includegraphics[width=1.0\columnwidth]{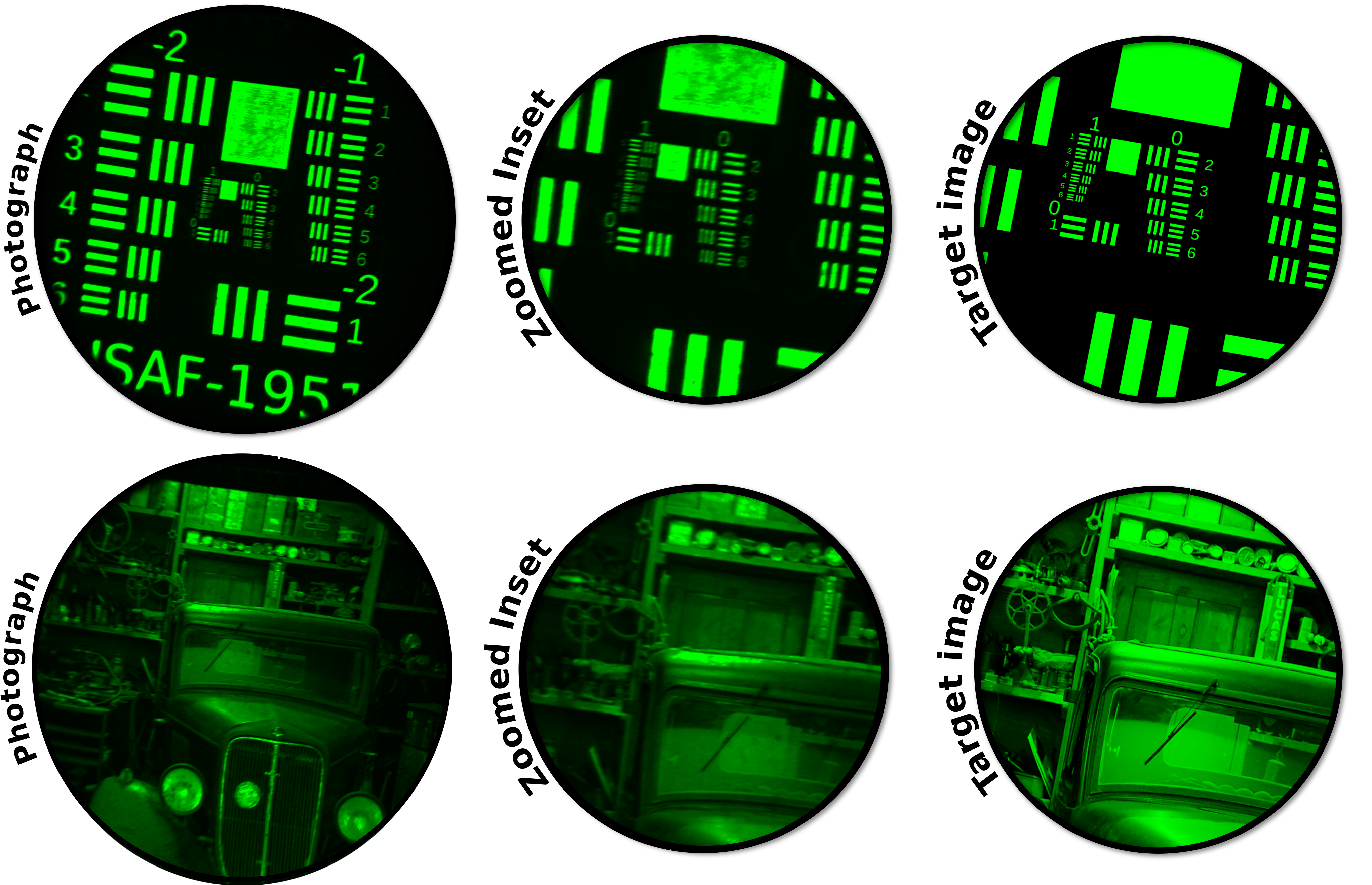}\vspace{-4mm}
\caption{
Captured photographs from the eyebox of our phase-only \projectname prototype.
These photographs are captured using a XIMEA MC245MG-SY-UB image sensor equipped with an adjustable $5-50$ mm lens while using $20 ms$ exposure times.
During these captures, only a green LED light source is used in the display prototype.
The figure also shows zoomed-insets and their target images.
The provided images are circular as the aperture of our eyepiece in this prototype is circular.
The source image on the right side is from DIV2K \cite{agustsson2017ntire}.
}
\label{fig:image_quality}
\vspace{-4mm}
\end{figure}

\subsection{\projectname Software}
We choose to dedicate our amplitude-only prototype to showcasing thin eyepieces.
Thus, this particular prototype relies on projecting conventional 2D images to deliver its message related to form factor (not amplitude-only holograms).
In the meantime, we dedicate our phase-only prototype to delivering \3D images while providing a wide \FoV and large throw distances.
We choose not to use \HOE in the phase-only prototype to avoid the heavy engineering work that poses an engineering resource challenge (\eg, recording \HOE in multicolor, replicating optics of the display for the reference beam of a recording setup).

Our learned \CGH pipeline is simply dedicated to our phase-only prototype, and it bases on PyTorch \cite{paszke2017automatic} and a \CGH toolkit \cite{aksit_kaan_2022_6528486, kavakli2022introduction, kavakli2022optimizing}.
Training of our learned \CGH pipeline is conducted using a learning rate of 0.0001 for ten epochs (Source code: \codebase \cite{multiholo2022}).
Our model relied on a U-Net \cite{ronneberger2015u} with 28 hidden channels as our neural network.
To generate our dataset for training our learned \CGH pipeline, we first create depth maps for the DIV2K image dataset \cite{Agustsson_2017_CVPR_Workshops} following the work by Alhashim and Wonka\etal \cite{alhashim2018high, densedepth2022}.
We resize these images and their depth to the resolution of our \SLM.
Using their estimated depth, we use these images to generate multiplane holograms with six planes following the work by Akşit \etal \cite{relisticdefocus2022,kavakli2022realistic} (plane separation is 1 mm).
We discard the estimated depth, and we use 900 input images and their corresponding multiplane holograms in our training while 100 of them are in validation.
As the training complete, our estimation routine takes about 20-28 ms to generate a single phase-only hologram at the resolution of our \SLM.
We use a computer with NVIDIA GeForce RTX 2080 GPU with 12~GB memory and an Intel i7, 3.9 GHz CPU to drive our holographic display prototype.
When we display our holograms in our phase-only prototype, we update the calculated $O_h$ with a linear phase grating term to avoid undiffracted light,
\begin{equation}
  O_h'(x,y) =
  \begin{cases}
            e^{-j(\phi(x,y) + \pi)} & \text{if $y=$ odd} \\
            e^{-j\phi(x,y)}         & \text{if $y=$ even}, \\
  \end{cases}
\end{equation}
where $\phi$, $x$, $y$ represents the original phase of $O_h$.
In Listing \ref{list:forward_model}, we provide a simplified learned estimation model that used in our training and estimation routines.
Note that the estimation routine relies on a \CNN. 
Specifically, we use a U-Net as our \CNN \cite{ronneberger2015u}.
This \CNN takes a single color of an image as input.
The output of the \CNN is a tensor with two channels.
The resultant constrained tensor represents the phase component of a phase-only hologram.
During a training session, the output of this estimation model is compared against a ground-truth hologram using an L2 loss function.
Unlike the recent literature \cite{choi2021neural}, no reconstruction losses are involved.

\section{Evaluation}
Using our two \projectname prototypes, we assess the practical limits of our approach.
For analyzing the optical qualities of our \projectname approach, we use our phase-only prototype, whereas for the demonstration of a slim eyepiece built using an in-house recorded \HOE, we will use our amplitude-only prototype.

\subsection{Quality Analysis}
In this section, we will rely on optimized phase-only holograms for assessing absolute resolution and \FoV characteristics of our proposed method.
For demonstrating, \3D images from our prototype, we will rely on both our learned method and optimization method.

\begin{figure}[tb]
\includegraphics[width=0.9\columnwidth]{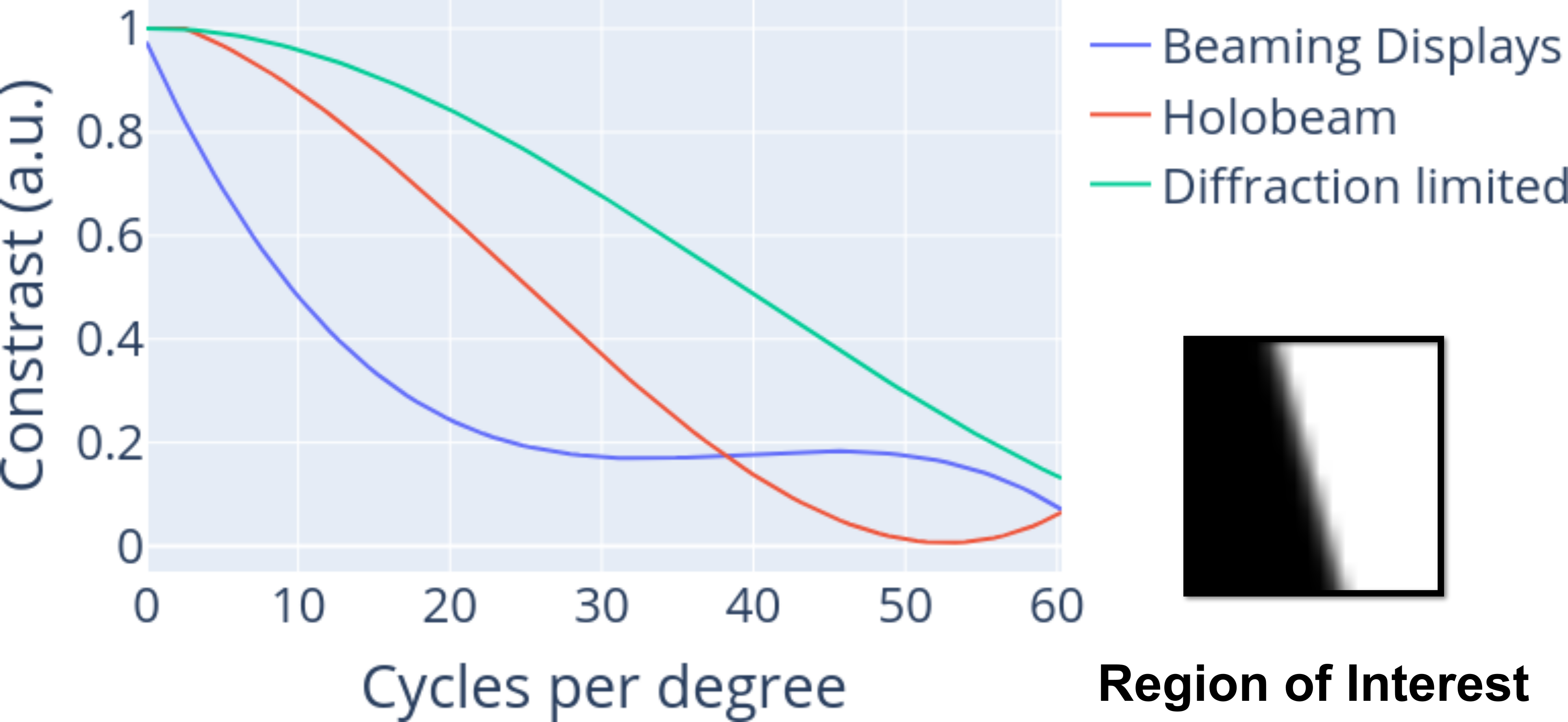}\vspace{-3mm}
\caption{
Modulation Transfer Function Analysis for phase-only \projectname prototype with LED illumination.
Beaming Displays \cite{itoh2021beaming} provided 8 \cpd at their central \FoV.
Phase-only \projectname prototype supports 24 \cpd at half contrast in its central \FoV.
Thus, \projectname improves the resolution three folds while expanding the \FoV with respect to Beaming Displays \cite{itoh2021beaming}.
}
\label{fig:modulation_transfer_function}
\vspace{-4mm}
\end{figure}

\paragraph{Resolution and Field of View.}
The eye relief of our phase-only prototype is $35$ mm.
The aperture of our eyepiece is $50.8$ mm, leading to \FoV of $70$ degrees as depicted in \refFig{teaser}.
We also provide additional results from various scenes in \refFig{image_quality}.
To assess the resolution quality of this prototype, we rely on a standard \MTF analysis \cite{burns2000slanted}.
Across our evaluations, we capture photographs with a XIMEA MC245MG-SY-UB image sensor and an adjustable $5-50$ mm lens while using $20$ ms exposure times to approximate a human observer's experience.
Our \MTF analysis suggests that the phase-only \projectname prototype can support up to $24$ cpd in resolution when used with LED illuminations (at central \FoV).
A healthy \HVS demands $30$ \cpd or more for realistic-looking resolutions.
Such resolutions could be met as we use the existing lasers in our phase-only \projectname prototype.
However, we observe that perturbating fringe patterns shadow the image quality as depicted with an extra capture at our supplementary.
We believe these perturbations mostly originated from optics starting from the end of 4f lenses towards the eyepiece.
\projectname could potentially provide higher resolutions while using LEDs with the increasing aperture size of \SLM and lenses in the future.

\paragraph{Multiplane Images.}
Light source coherency used in a holographic display dictates the depth of field of the reconstructed images \cite{lee2020light}.
Thus, incoherent broadband sources increase the depth of field and degrade optical focus cues in reconstructed images.
Given the situation with the depth of field of images, the conventional way to demonstrate multiple images in a holographic display involves using a laser light source.
We also use laser in our phase-only \projectname prototype to demonstrate capabilities related to multiplane images.

\begin{figure}[tb]
\includegraphics[width=1.0\columnwidth]{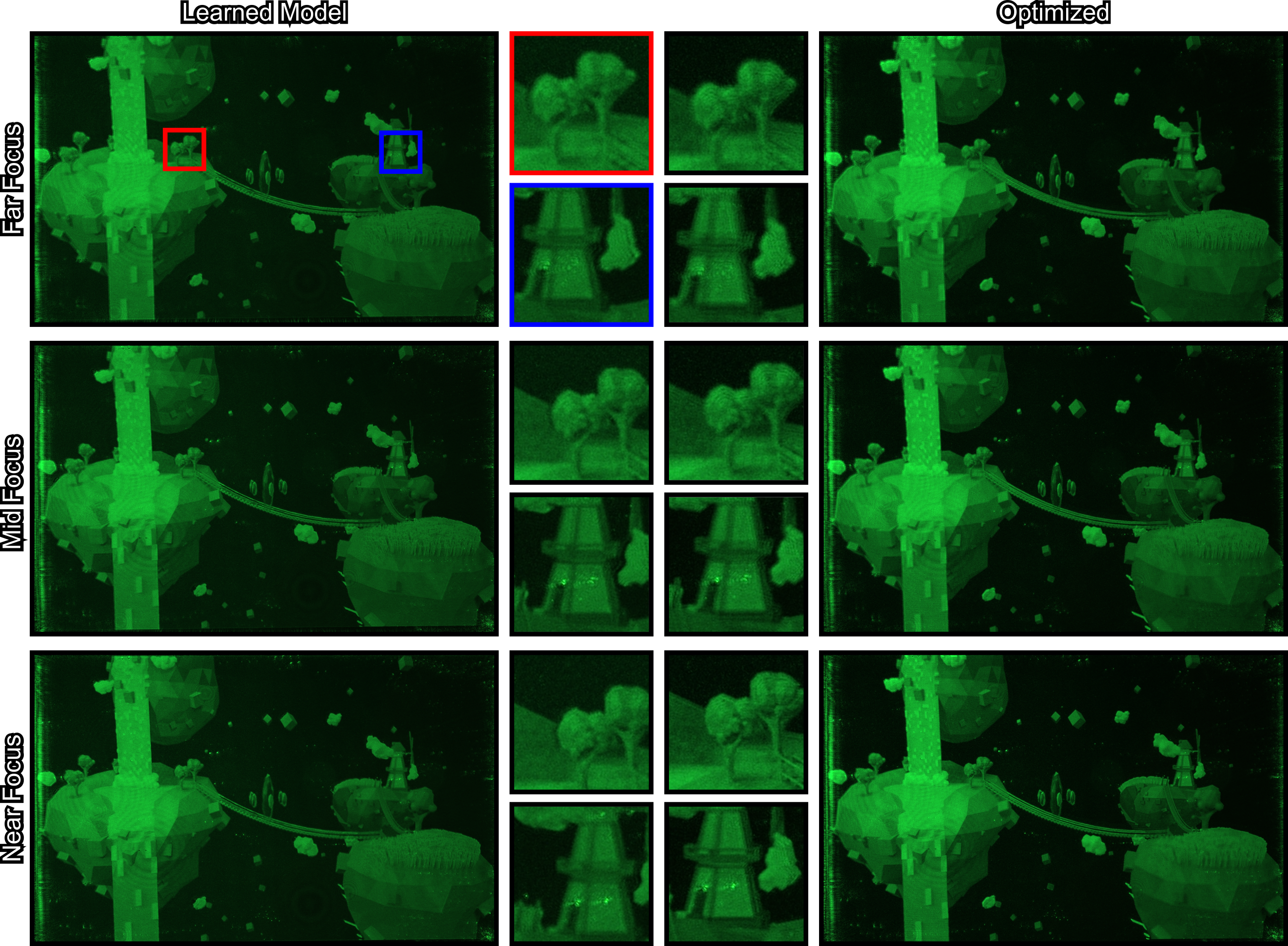}\vspace{-3mm}
\caption{Learned Model for Multiplane Holograms. 
Our learned hologram generation method can help reconstruct high-quality images with depth support from conventional images.
However, the generated hologram's depth is not necessarily faithful to the original depth as compared with the optimized hologram.
We observe that the learned model tends to shrink the depth range (\eg, six multiplane images in this example are mostly focused on one to three planes).
Note that the optimized case resembles a typical outcome from recent standard literature \cite{kavakli2022realistic,choi2021neural,shi2022end}.
For more results, consult our supplementary.
}
\label{fig:multiplane}
\vspace{-5mm}
\end{figure}

However, as the lenses between our 4f system and eyepiece generate unintended aberrations and distortions, our final image contains fringes when lasers are used (see our supplementary document for evidence).
These distortions could be fixed in the future using learned approaches that could account for imperfections in holographic display hardware \cite{kavakli2022learned,choi2021neural,chakravarthula2020learned} and by designing and manufacturing dedicated projection optics like in many projector products.
Thus, to avoid any visual artifacts caused by the eyepiece and additional lenses, we capture reconstructions of our phase-only holograms right after the 4f imaging system using a bare XIMEA MC245MG-SY-UB image sensor with $50 ms$ exposure.
We provide a sample capture demonstrating a focus change as in \refFig{multiplane}. 
Note that the optimized version shown in \refFig{multiplane} resembles the image quality of most recent standard literature \cite{kavakli2022realistic,choi2021neural,shi2022end}.
For more results, please consult our supplementary materials.

In our observation, the learned model shrinks the depth range of a scene (\eg, six multiplane images mostly focused at two planes) and tends to distribute depth levels that are not faithful to an original depth map (\eg, windmills door becoming sharp at near focus rather than far focus in \refFig{multiplane}).
We believe this learned model promises encouraging first results towards a hologram generation routine where a 2D image is transformed into a 3D multiplanar hologram without requiring scene-depth information or multiple perspective images.

\subsection{HOE Lens Analysis}\label{subsec:hoe_analysis}
In \projectname, the spatial relationships among the projection system, the eyepiece, and the eye affect the final image quality and visibility. This section analyzes our \HOE design with simulations that mimics our amplitude-only prototype.

\begin{figure}[tb]
\includegraphics[width=1.0\columnwidth]{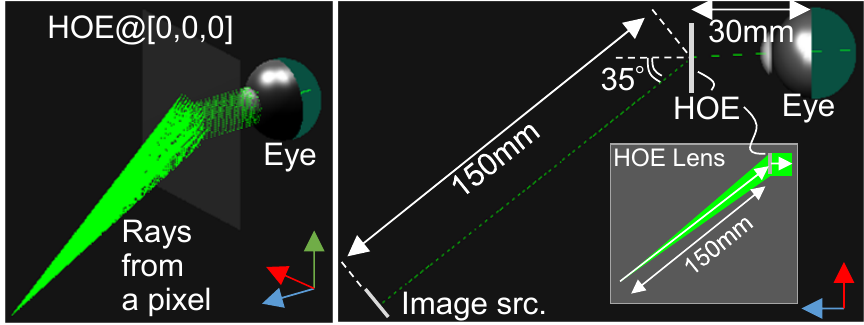}\vspace{-3mm}
\caption{
Visualization of our simulation in Sec.~\ref{subsec:hoe_analysis}. 
The configuration shows the default layout of each component. 
We used left-handed coordinates where the z-axis faces away from the viewpoint.
}
\label{fig:simulation_setup}
\vspace{-5mm}
\end{figure}

\begin{figure*}[tb]
\centering
\includegraphics[width=0.9\textwidth]{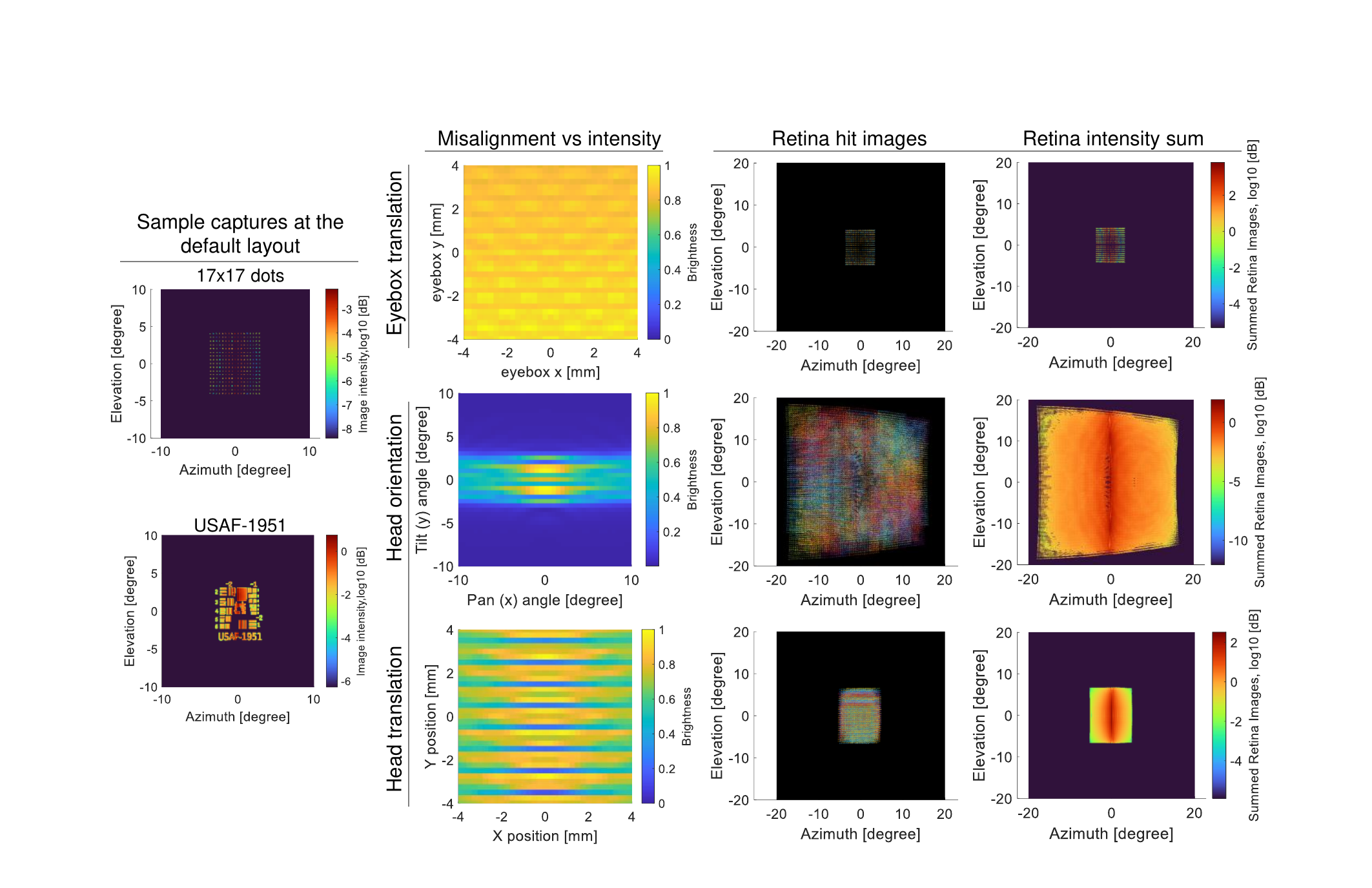}\vspace{-3mm}
\caption{A summary of the simulation analysis in Sec.~\ref{subsec:hoe_analysis}. The first column is simulated retinal views on the default layout when displaying a 17x17 point grid used in the analysis and a USAF1951 chart as another example. For the column between the second and fourth, the three rows for the eyebox, the head orientation, and the head translation analysis, respectively. The three columns visualize the relative image intensity over misalignment parameter spaces, visualization of the retinal hit count of the rays at sampled viewpoints, and of accumulated intensities.}
\label{fig:simulation_analysis}
\vspace{-5mm}
\end{figure*}

\paragraph{Simulation Setup}
In our simulation on MATLAB 2022a, we use an open-source ray optics library.
Since this library does not implement HOE simulations, we implemented a volume hologram class based on Sec.~\ref{subsec:eyepiece}.
For all simulations, we set the wavelength to $\wavelength=532$ nm, the average refraction index to $\refractiveIndexAverage=1.5$, the maximum refractive index modulation to $\refractiveIndexModulation=0.04$, and the photopolymer thickness $=30\mathrm{\mu m}$, respectively.

Figure~\ref{fig:simulation_setup} shows the layout of our simulation.
We used the library's default eye optics parameters for the eye model, and the pupil diameter was set to 3 mm.
The eyeball center is set to 30 mm from the \HOE lens.
We assume that the projection optics relays a virtual image from the amplitude-only \SLM and the virtual image forms at 150 mm from the HOE with a pan angle of 35 degrees.
The HOE is designed to form a lens with f=150 mm and a 35-degree angle tilt.
Note that we scaled the virtual image size by 3, assuming that the projection optics relay the \SLM image plane, leading to the pixel pitch of the relayed image as $\pixelpitch=5.4*3=16.2 \mu$m.
This results in the rays' diffraction angle from each pixel becoming $2\asin(1.22\wavelength / \pixelpitch)$, about 4.6 degrees.
We randomly generated 100 rays for each pixel along the central ray direction within the diffraction angle.

We also assume an ideal projection system that can guide images to the default pupil center position. We thus let the central rays target the center of the HOE, i.e. (0,0,0) in the world coordinates.
We refer to this positional relationship as the default layout. As a displayed image, we set 17x17-point grids uniformly spanning in an 801x801 pixel image (\refFig{simulation_analysis} left column top).

We evaluate two major factors related to misalignments: the eyebox and head alignment.
The eyebox analysis evaluates when the eye locations change while the virtual image source and the eyepiece are fixed.
The head alignment analysis evaluates when the head system, i.e., the HOE eyepiece and the eye, changes its orientations and positions against the virtual image source.

We set the disturbance ranges of each setup in the following.
For the eyebox position: $\pm$4 mm with 0.25 mm step in the x-y directions; for the head orientation: $\pm$10 degrees with 0.5-degree step for the pan-tilt angles; and for the head translation: $\pm$4 mm with 0.25 mm step in the x-y directions.

For the eyebox and head translation analyses, one may also explore the z direction.
We, however, fixed the z-axis parameter in our analyses to keep the explanations concise since our pilot analysis results did not give much difference in the z-axis analysis compared to the x-y space analyses.

\paragraph{Simulation Results}
Figure~\ref{fig:simulation_analysis} shows an overview of our simulations, where the second to fourth rows correspond to eyebox analysis (x-y plane), head orientation analysis (pan-tilt angles), and head translation analysis (x-y plane), respectively.
The second column shows the total brightness of the observed images for the given misalignment parameters.
The total brightness is relative to that of the image taken with the default layout, i.e., the center pixel of each figure.
The third column is a color visualization of whether or not the light rays hit the retina at each viewpoint.
The images accumulated hit from every 10 viewpoints over the misalignment parameter space.
And the rays from the same viewpoint will be the same color regardless of their intensity.
The fourth columns instead accumulate ray brightness and show colormaps of the total brightness of all given view conditions.
Please also refer to our supplementary viewpoint videos of the three analyses.

The eyebox analysis shows that the brightness level is consistent over eye position changes.
This is understandable given the diffraction angle of the virtual image is about 4.6 degrees, which sufficiently covers the pupil, as we can also observe in \refFig{simulation_setup}.

The head orientation analysis shows that our setup is more robust in the pan (horizontal) rotation than in the tilt (vertical) orientation.
A possible reason for this tendency is that the HOE lens is designed for input rays' virtual point source to be placed on the x-z plane; thus, the diffraction efficiency could radically decrease (off-Bragg) along vertical angle errors.

Finally, the head translation analysis shows that our setup is more robust in head translation errors than head orientation errors.
However, the image intensities change periodically along vertical head misalignment.
The image thus may appear to flicker in dynamic tracking environments, depending on the tracking accuracy.
This tendency may stem from the periodical structure of the Bragg diffraction efficiency over input angles, as in \refFig{diffraction_examples}.
Our informal, practical observation with the amplitude-only prototype aligns with the simulation results.

\section{Discussion and Limitations}
\label{sec:discussion_and_limitations}
Beaming Displays \cite{itoh2021beaming} design methodology aims to resolve a vital issue in \AR glasses designs and enable practical \AR applications that could converge to comfortable and realistic \AR experiences.
Although \projectname provided evidence that Beaming Display designs could provide resolutions and \FoV meeting more of \HVS demands, there are still outstanding major issues related to Beaming Displays design methodology.
We list these issues and their potential solutions in the following paragraphs.

\paragraph{Eyebox and cost}
Holographic approaches imply small eyebox traditionally, and also true for our implementation.
Most recently, the work by Jang \etal \cite{jang2022waveguide} addresses the eyebox issue while maintaining a true \3D holographic \AR glasses, which could help inspire a fix of the issue.
Holographic displays lead to a complex hardware design that is highly costly due to niche equipment such as a phase-only \SLM \cite{zhang2014fundamentals}.
Our amplitude-only prototype aims to improve on cost of holographic displays by relying on cheaper but less capable hardware.
However, this is not a true fix to the cost problem as cheaper alternatives for a phase-only \SLM have to be invented.

\paragraph{Improving mobility.}
Beaming Displays \cite{itoh2021beaming} provided evidence that delivering images to a moving user could be possible to some extent, however, \projectname does not provide any tracking of users and the same case is not replicated as \projectname focused on improving image quality and form factor.
The original Beaming Displays work tends to relay an image on a screen integrated into an eyepiece, whereas \projectname is a standalone optical relay that magnifies im- ages projected in mid-air. 
Thus, \projectname is much less forgiving of any misalignment in the beaming path. 
We believe this requires special attention in future works and the functionality of a diffuser in the original Beaming Displays has to be embedded into future variants \projectname to benefit from both mobility and image quality improvements.

If the functionality of a diffuser could be transferred to a new eyepiece design, we trust that mobility-related issues (e.g., head position and orientation of a user) could be resolved without sacrificing the form factor.
This item is our next step in this line of research.

\paragraph{Improving the eyepiece.}
The image brightness provided by a \HOE eyepiece is sensitive to misalignments due to users' head orientations.
We believe there are potential research directions that could help mitigate these alignment-related issues.
A potential solution could be to investigate embedding the trends in curved or free-form \HOE designs \cite{bang2019curved,jang2020design} to our \HOE designs in the future.
Another potential research direction is to bring the properties of bird-bath optics used in the original Beaming Displays \cite{itoh2021beaming} into the \HOE design space.
In such a case, we could explore the embedding properties of a diffuser with a lens in \HOE designs~\cite{yeom2021projection, hwang2022uniformity,yeom2022analysis}.

\paragraph{\projectname's target applications.}
Given the limitations in tracking and alignment for \projectname, we identify applications that could make the most sense for what is possible with \projectname today.
Applications that do not require a user to be frequently moving could be more forgiving with the lack of tracking.
Thus, we believe \projectname could be a good fit for work-related applications (\eg, desktop displays).
There are existing \3D desktop displays for similar purposes 
(\eg, Brelyon\footnote{https://brelyon.com/}), 
which we believe \projectname could provide a new slim version.
A potential application area in the future could also be automotive displays, namely heads-up displays.
Note that a driver in a car is more or less stationary and must constantly gaze at the road.
A future variant of \projectname could provide more freedom in movement and could be a potential new way to build a heads-up display that does not require fiddling with the windshield or dashboard of a car.

\paragraph{Improving Depth Generation of Learned \CGH method.}
The depth generation of our learned holograms is limited as the depth information of a scene is not provided as an input to our learned algorithm.
We designed our learned algorithm as a single constrained U-Net.
However, a depth estimation network could also be used to condition our method or a more detailed analysis could be conducted in the future regarding the contribution of each layer in that single U-Net.
This way, an educated way to improve our attempts could be achieved in the future.

\paragraph{Complementing contact lens \AR.}
In recent years, there has been a push in research related to \AR contact lenses \cite{sano2021holographic}, which could potentially transform \AR to a seamless setting without any glasses.
In the future, our work could also be transformed into an optical component embedded in a contact lens that helps register beams from a projector to a user's retina.

\section{Conclusion}
\label{sec:conclusion}
Future AR glasses must be thin and lightweight while supporting high resolutions, wide \FoV, and optical focus cues. 
Today’s AR glasses struggle to balance these requirements, often yielding to design challenges. 
However, the experts largely agree that these requirements are compulsory to maintain comfortable visual experiences with AR glasses. 
So, we ask whether the way out of this struggle could lie in changing the perspectives in AR glasses design.

This paper introduces a new stand leading to a bold change in AR glasses design which we named \projectname.
For the first time in the literature, we demonstrate very slim AR glasses based on our \projectname method.
These AR glasses support \3D images with near-correct optical focus cues and provide resolutions matching commonly accepted retinal resolution criteria (30 cpd).
In this design, there are still open research issues that require proper attention from the relevant research communities.
These issues include tracking the accuracy of users beaming accuracy of projectors and designs, leading to more freedom in user movement.
Although these outstanding issues exist, this work is a significant milestone that could pave the road to a new body of work toward the ultimate AR glasses with comfortable and realistic visual experiences.

\section*{Supplementary material}
Our code  is available at \codebase.
Dependencies are available at \dependbase~\cite{kavakli2022optimizing, kavakli2022introduction}.
Additional media, results and materials are distributed separately as a supplementary documents.

\acknowledgments{
The authors wish to thank Koray Kavaklı for fruitful discussions.
Kaan Ak\c{s}it is supported by the Royal Society's RGS\textbackslash R2\textbackslash 212229 - Research Grants 2021 Round 2 in building the hardware prototype and Meta Reality Labs inclusive rendering initiative for building the rendering pipeline.
Yuta Itoh is supported by JST FOREST Grant Number JPMJFR206E and JSPS KAKENHI Grant Number JP20J14971, 20H05958, and 21K19788, Japan.
}

\bibliographystyle{abbrv-doi}
\bibliography{references}

\AtEndDocument{

\section*{Supplementary material}

# Supplementary Materials for "HoloBeam: Paper-Thin Near-Eye Displays"

Kaan Akşit*
University College London

Yuta Itoh†
The University of Tokyo

## 1 CONTRIBUTIONS FROM EACH AUTHOR

Here we list contributions from each author in this specific research work:

Yuta Itoh:

- Designing and building a HOE recording setup in hardware.
- Compiling extensive study on HOE design space.
- Compiling figures for HOE recording setup and functional test of recorded HOEs.
- Review and edits for the entire documentation.

Kaan Akşit:

- Initiation of project idea of *HoloBeam*, this includes blueprints for projector designs, conventional eyepiece designs and algoritmic designs.
- Design and manufacturing of amplitude-only and phase-only *HoloBeam* prototypes,
- Coding and training of the machine learning model.
- Writing abstract, introduction, related work, conclusion sections entirely.
- Compiling figures for a simplified layout diagram of *HoloBeam*, teaser, phase-only and amplitude-only hardware in the manuscript and supplementary documentation.
- Review and edits for the entire documentation.

## 2 ADDITIONAL RELATED WORKS

### 2.1 Thin Relay Optics for Augmented Reality Glasses

**Holographic Optical Elements** There are various photosensitive materials for holograms [14]. For example, Silver halide emulsions are used for amplitude modulating Holographic Optical Element (HOE)s. Yet another example, Photorefractive crystals (dichromate gelatin or lithium niobate) and photopolymers are organic materials used for phase modulating HOEs. Among these materials, photopolymers are relatively easy to handle because they do not require the development process required for other materials and can be bleached by incoherent light. Attracting attention as a replacement for HOEs, metasurfaces are artificial sub-wavelength subwavelength structures promising superior optical performance, for example, large aperture, wide bandwidth, and polarization dependence [9, 10]. However, these systems are in their early stages in proving their practicality in Augmented Reality (AR) glasses.

---

*e-mail: k.aksit@ucl.ac.uk
†e-mail: yuta.itoh@iii.u-tokyo.ac.jp

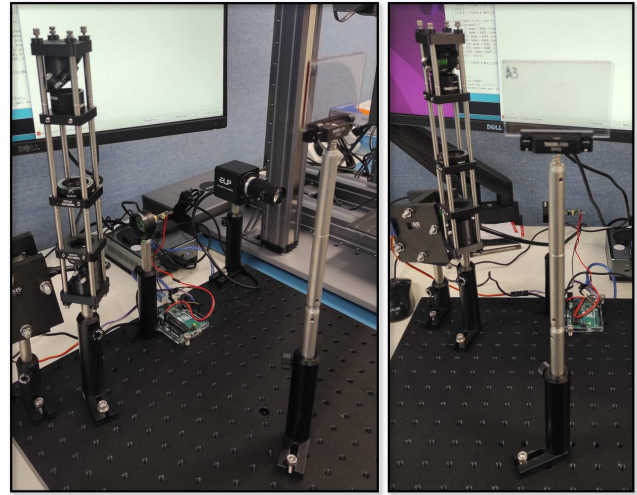

Figure 1: Vertically built amplitude-only *HoloBeam* prototype. Our projection assembly's optical path is folded using two mirrors in this configuration.

### 2.2 Augmented Reality Glasses

**Beaming AR Glasses.** Most recently, a new design method called Beaming Displays [6] is proposed, where active components are separated from AR glasses designs to address issues in conventional AR glasses providing a fixed focus image. In their work, a user wears bird-bath optics similar to the ones in the literature, and the image is projected into this bird-bath optics from a distance. However, their results [6] demonstrate a moderate Field Of View (FoV), limited resolutions, and bulky form factor with no support in optical focus cues. Our work follows the design approach in Beaming Displays in AR glasses. However, *HoloBeam* merges the benefits of holographic techniques with Beaming Displays. This way, *HoloBeam* improves upon Beaming Displays by supporting optical focus cues, improved resolutions and unmatched form-factors.

**Other AR apporaches.** We do not cover all other AR displays approaches in the literature that may have a loose relevance to our work [3]. Both previous literature [6] and *HoloBeam* share similarities with projection based spatial AR displays such as head-up displays in automative [4]. The primary difference in our approach with projection based spatial AR displays is that the eyepiece is close to a user's eye. Although there are AR displays having eyepiece close to a user's eye [8], unlike theirs our solution do not require a large projection screen. Thus, *HoloBeam* brings the possibility to widen FoV and making displayed images visible only to a user wearing our eyepiece. In recent years, there has been a push in research related to AR contact lenses [12], which could potentially transform AR to a seamless setting without any glasses. We understand that such contact lenses based approaches pose primary challenges in practical aspects and approvals. In the future, our work could also be transformed into an optical component embedded in a contact lens

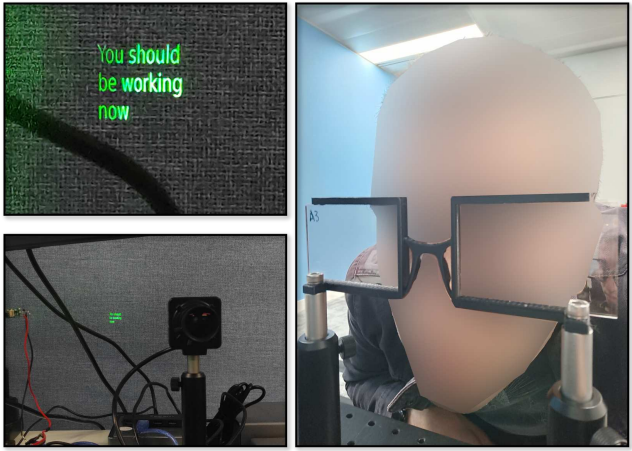

Figure 2: Various photographs showing usage and view from the HOE eyepiece. Note that this eyepiece is dedicated to amplitude-only *HoloBeam* prototype.

that helps register beams from a projector to a user's retina. Thus, *HoloBeam* has the potential to be complementary to the contact lens based AR designs of the future.

## 3 AMPLITUDE-ONLY *HoloBeam* PROTOTYPE

Our amplitude-only *HoloBeam* prototype could also be built in a vertical layout as demonstrated in Fig. 1. This way, we demonstrate that our projection assembly could be assembled in various forms using optical path folding (e.g., with mirrors or prisms). Such path folding could be helpful when integrating *HoloBeam* to an automative application in a potential future.

It should also be noted that we provide sample photographs from our benchtop prototype in Fig. 2.

## 4 PHASE-ONLY *HoloBeam* PROTOTYPE

Additional photographs from our phase-only *HoloBeam* prototype are provided in Fig. 3.

## 5 OPTICAL BEAM PROPAGATION

A phase-only hologram is described as a two-dimensional array filled with phase values and typically described with a complex notation as $O_h = e^{j\phi(x,y)}$, where $\phi$ represents the phase delay introduced by each pixel at a phase-only hologram. Holographic displays typically represent holograms, $O_h$ with programmable SLMs. Meanwhile, a coherent beam illuminating a phase-only hologram, $U_i$, is also described as a two-dimensional array. Note that $U_i$ is an oscillating electric field described as $U_i = A_0 e^{j(k\vec{r}+\phi_0(x,y))}$, where $A_0$ represents the amplitude of the optical beam, $k$ represents the wavenumber that can be calculated as $\frac{2\pi}{\lambda}$, $\lambda$ represents the wavelength of light, and $\phi_0$ represents the initial phase of the optical beam. In calculation, $A_0$ is often considered $A_0 = 1$ for an ideal collimated beam, while $\phi_0$ is assumed to be a two-dimensional array filled with random values between zero to $2\pi$. Finally, leading to simplification of $U_i$ as $e^{j\phi_0}$. In simple terms, as $U_i$ illuminates $O_h$, $U_i$ by modulated with $O_h$, forming a new modulated beam $U_m$ that is calculated as

$$U_m = U_i O_h = e^{j(\phi(x,y)+\phi_0(x,y))}. \quad (1)$$

A modulated beam, $U_m$, has to propagate in free-space from the hologram plane (SLM plane) towards a target depth plane to reconstruct images at various depth planes. Propagation of optical beams from one plane to another follows the theory and method introduced by Rayleigh-Sommerfeld diffraction integrals [5]. This diffraction integral's first solution, the Huygens-Fresnel principle, is expressed as follows:

$$u(x,y) = \frac{1}{j\lambda} \int \int u_0(x,y) \frac{e^{jkr}}{r} cos(\theta) dx dy, \quad (2)$$

where resultant field, $U(x,y)$ is calculated by integrating over every point across hologram plane, $U_0(x,y)$ represents the optical field in the hologram plane for every point across XY axes, $r$ represents the optical path between a selected point in hologram plane and a selected point in target plane, $\theta$ represents the angle between these points. The angular spectrum method, an approximation of the Huygens-Fresnel principle, is often simplified into a single convolution with a fixed spatially invariant complex kernel, $h(x,y)$ [13],

$$u(x,y) = u_0(x,y) * h(x,y) = \mathscr{F}^{-1}(\mathscr{F}(u_0(x,y))\mathscr{F}(h(x,y))). \quad (3)$$

In our implementations, we rely on a fundamental library for optical sciences [2], which provides a differentiable version of various optical beam propagation methods. Therefore, our methods can work with other beam propagation approximations.

## 6 ADDITIONAL RESULTS FROM PHASE-ONLY *HoloBeam* PROTOTYPE

In this section, we provide additional results from the eyebox of our prototype. Fig. 4 depicts the effect of using coherent and incoherent illumination sources in the phase-only *HoloBeam* prototype.

Fig. 5 demonstrates a full color image reconstruction using our phase-only *HoloBeam* prototype. Please note that this demonstrator does not have color distortion correction implemented, yet. Therefore, in Fig. 5, readers will observe color distortions and misalignments in the final image.

Fig. 6 provides additional captures from our phase-only *HoloBeam* prototype, demonstrating the image quality.

For more results from our prototypes, please consult to provided compressed files that contains videos and raw photographs from our experiments.

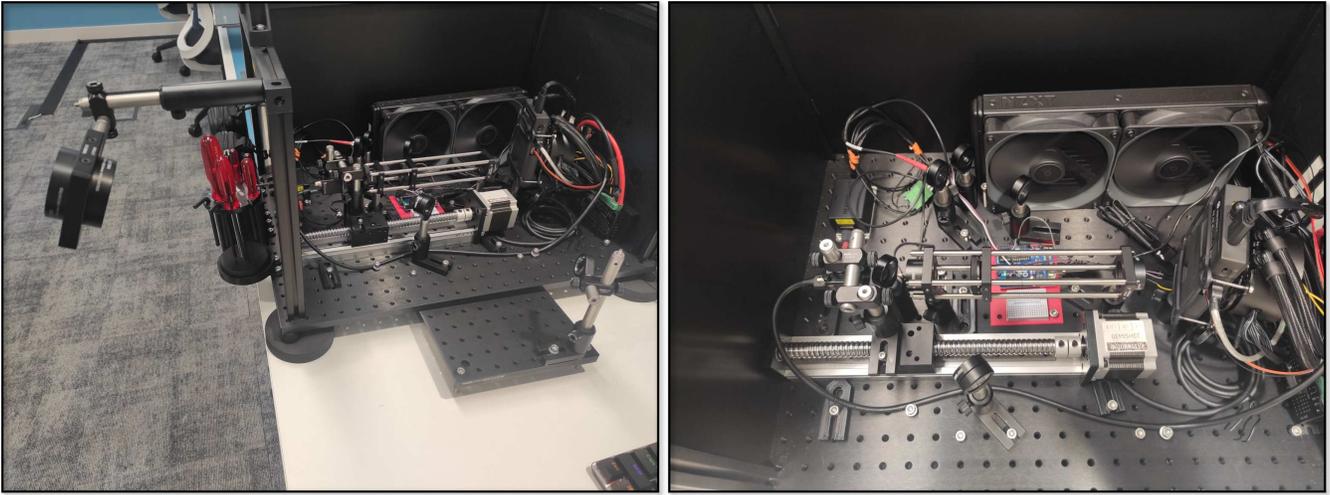

Figure 3: Additional photographs from our phase-only *HoloBeam* prototype. Images shows a true scale assembly and close look up at the projector assembly. Cooler fan in this hardware setup helps keeping the Phase SLM stable at a fix temperature. In a final product, these components are not necessarily needed, and there are much custom compact holographic projectors in some existing products (e.g., Holoeye).

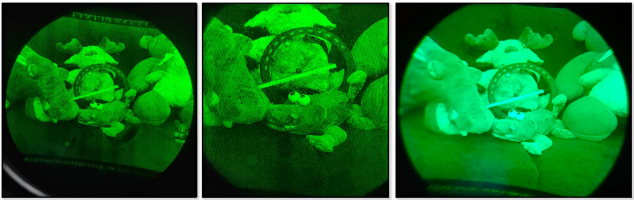

Figure 4: Photographs from the eyebox of the phase-only *HoloBeam* prototype. (Left) A photograph showing the resultant reconstructed image when a single color of the laser is used. (Middle) A close-up view from the very same scene shown in the left image. (Right) A photigraph showing the resultant reconstructed image when a single color LED is used. Lasers cause unintended perturbations in the image as demonstrated in this figure, while LEDs lead to a smoother image. Note that these photographs were taken using a smartphone camera.

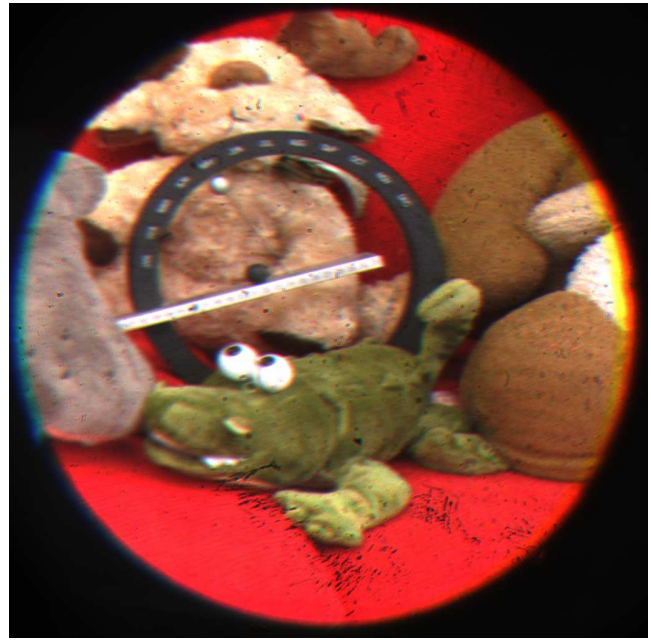

Figure 5: Photographs from the phase-only *HoloBeam* prototype. Full color photographs from our *HoloBeam* prototype. Note that full color images do not have the correct distortion correction in-place, thus leading to aberated images.

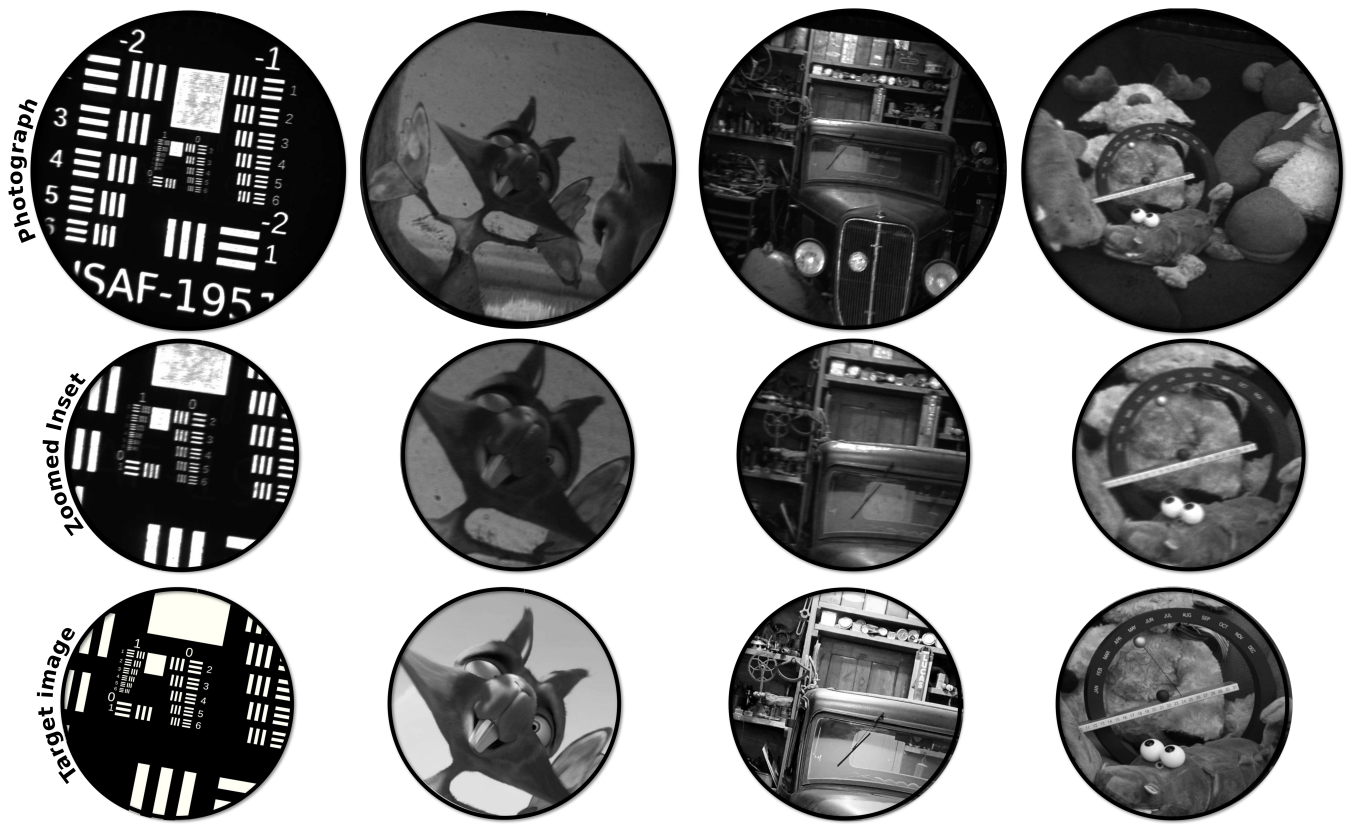

Figure 6: Resolution Quality of phase-only *HoloBeam* prototype. The first row shows captured photographs from the eyebox of our phase-only *HoloBeam* prototype. These photographs are captured using a XIMEA MC245MG-SY-UB image sensor equipped with an adjustable $5-50$ mm lens while using $20ms$ exposure times to approximate a human observer's view. During these captures, only a green LED light source is used in the display prototype. The second row shows zoomed-insets from the first row, while the third row shows the target images. The provided images are circular as the aperture of our eyepiece in this prototype is circular. The image quality of our phase-only *HoloBeam* approximates the ground truth images with a slight loss in contrast. The source image in the second, third and fourth column are from Big Buck Bunny [11], DIV2K [1] and Changil et al. [7], respectively.

}
\end{document}